\documentclass[reqno, 12pt]{amsart}
\usepackage[ansinew]{inputenc}
\usepackage{amsfonts}
\usepackage{latexsym}
\usepackage{amsmath}
\usepackage{graphicx}
\usepackage{color}

\newtheorem{defin}{Definition}[section]

\begin{document}

\footnote{*Corresponding author: Jian-Guo Liu, E-mail address:
395625298@qq.com. Project supported by National Natural Science Foundation of
China (Grant No 81860771).}

\begin{center}{\bf Stripe solitons and lump solutions to a generalized
(3 + 1)-dimensional B-type Kadomtsev-Petviashvili equation with
variable coefficients in fluid dynamics}
\end{center}

\vskip 4mm

\begin{center} Wen-Hui Zhu $^{1}$, Jian-Guo Liu$^{2,*}$,
\end{center} \vskip 2mm
\noindent {1} Institute of artificial intelligence, Nanchang
Institute of Science and Technology, Jiangxi 330108, China\\
\noindent {2} College of Computer, Jiangxi University of
Traditional Chinese Medicine, Jiangxi 330004, China
 \vskip 6mm

{\bf Abstract:}\,Under investigation is a generalized (3 +
1)-dimensional B-type Kadomtsev-Petviashvili equation with variable
coefficients in fluid dynamics. Based on the Hirota's bilinear form
and the positive quadratic function,
 abundant lump solutions are obtained. The interaction solutions between lump solutions and other solitons are also presented.
 Their dynamical behaviors are graphically shown with different choices of the
free parameters.\vskip 2mm
 {\bf Keywords:}\,Kadomtsev-Petviashvili equation; variable
coefficients; dynamical behaviors; lump solutions \vskip 2mm

\vskip 4mm

\noindent {\bf 1. Introduction}\\

\quad Lump is a kind of rational solution with large amplitude. Searching for the lump solution is one of the hot issues. In 2015,  a direct and efficient algebraic method to obtain the lump solution is proposed in Ref. [1]. Subsequently, the theoretical support for this method is provided and proved (see Theorem 1.1 [2]), which made the method was widely used for deriving the lump and interaction solutions of integrable equations, such as general combined fourth-order soliton equation [3],  combined fourth-order equation [4], new (3+1)-dimensional Hirota Bilinear equation [5], generalized Hietarinta-
type equation [6],  generalized Hirota-Satsuma-Ito equation [7], second KPI equation [8], (2+1)-dimensional Ito equation [9] et al.

\quad Kadomtsev-Petviashvili (KP) equation was a completely
integrable equation to represent nonlinear wave motion. The KP
equation can be applied to describe water waves of long wavelength
with weakly non-linear restoring forces and frequency dispersion
[10]. Particularly, the constant-coefficient KP equations have been
studied in many works, such as multi-soliton solutions [11],
conservation laws [12], Bilinear forms [13] and lump solutions [14-17].
 However, the variable-coefficient KP equations can furnish more realistic
models than the constant-coefficient KP equations [18-22]. So, a
generalized (3 + 1)-dimensional variable-coefficient B-type KP
equation is investigated as follows [22]
\begin{eqnarray}
\rho  a(t) \left(u_x u_{xy}+u_y u_{xx}\right)+a(t) u_{xxxy}+b(t)
   \left(u_{zz}+u_{xx}\right)+u_{zt}+u_{yt}+u_{xt}=0.
\end{eqnarray}Eq. (1) depicts the
propagation of nonlinear waves in fluid dynamics. $a(t)$ and $b(t)$
are real functions.
 $\rho$ is non-zero constant. The auto-B\"{a}cklund transformation, Bilinear form, multi-soliton
 solutions and shock-wave-type solutions for Eq. (1) have been presented. But, neither rational solutions
 nor lump solutions have been obtained [22,23].

\quad The organization of this paper is as follows: Section 2
presents the lump solutions for Eq. (1) based on Hirota's bilinear
form and symbolic computation [24-35]. Section 3 discusses the
interaction solutions between lump solution and the stripe soliton.
Section 4 obtains the interaction solutions between lump solution
and hyperbolic function. The dynamical behaviors for these obtained
solutions are demonstrated by some three-dimensional graphs and
contour plots. Section 5 makes a summary.\\

\noindent {\bf 2. Lump solutions}\\

Based on the transformation $u=\frac{6}{\rho}\,[ln\xi(x,y,z,t)]_x$
and Bell polynomials [22], the bilinear form for Eq. (1) is
presented as follows{\begin{eqnarray} [D_t D_x+a(t) D_x^3
D_y+D_t\,D_y+D_t\,D_z+b(t)(D_x^2+D_z^2)] \xi\cdot \xi=0.
\end{eqnarray}
}This is equivalent to: {\begin{eqnarray} &&3 a(t) \xi_{xy}
\xi_{xx}-3 a(t) \xi_x \xi_{xxy}+b(t) \left(-\xi_z^2-\xi_x^2+\xi
\xi_{zz}+\xi  \xi_{xx}\right)+\xi \xi_{zt}\nonumber\\&&-\xi_t
\xi_z-\xi_t \xi_y+\xi \xi_{yt}-\xi_t \xi_x+\xi  \xi_{xt}-a(t) \xi_y
\xi_{xxx}+a(t) \xi  \xi_{xxxy}=0.
\end{eqnarray}}Supposing that Eq. (1) has following rational and lump solutions
{\begin{eqnarray} \zeta&=&\vartheta _4(t)+x \vartheta _1+y \vartheta _2+z \vartheta _3,\nonumber\\
\varsigma&=&\vartheta _{9}(t)+x \vartheta _6+y \vartheta _7+z \vartheta _8,\nonumber\\
\xi&=& \zeta^2+\varsigma^2+\vartheta _{5}(t),
\end{eqnarray}}where $\vartheta_i (i=1,2,3)$ and  $\vartheta_i (i=6,7,8)$ are
undefined constants. $\vartheta_4(t)$, $\vartheta_{9}(t)$ and
$\vartheta_{5}(t)$ are unknown differentiable function. Substituting
Eq. (4) into Eq. (3), the values of above parameters are obtained as
follows {\begin{eqnarray}(I): \vartheta _2&=&\vartheta _7=0,
\vartheta_4(t)= \eta _1-\int_1^t \frac{\left(\vartheta
_1^2+\vartheta _3^2\right) b(t)}{\vartheta _1+\vartheta _3} \, dt,
\vartheta _{5}(t)=\vartheta _{5},\nonumber\\
\vartheta_{9}(t)&=&\eta _2-\int_1^t \frac{\left(\vartheta
_1^2+\vartheta _3^2\right) \vartheta _6 b(t)}{\vartheta _1
\left(\vartheta _1+\vartheta _3\right)} \, dt,
\vartheta_8=\frac{\vartheta _3 \vartheta _6}{\vartheta _1}
\end{eqnarray}}with  $\vartheta _1 \left(\vartheta _1+\vartheta _3\right)\neq 0$.
{\begin{eqnarray}(II): \vartheta _2&=&\vartheta _7=0,
\vartheta_8=\frac{\vartheta _3 \vartheta _6}{\vartheta
_1},\nonumber\\ \vartheta_{9}(t)&=&\eta _3-\int_1^t [\left(\vartheta
_1^2+\vartheta _3^2\right) \left(\vartheta _1^2+\vartheta
_6^2\right) b(t)+\left(\vartheta _1+\vartheta _3\right) [\vartheta
   _1^2 \vartheta _4'(t)]]/[\vartheta _1 \left(\vartheta _1+\vartheta _3\right) \vartheta _6] \, dt,  \nonumber\\ \vartheta_{5}(t)&=& \eta _4+\int_1^t [2 \left(\vartheta _1 \vartheta _{9}(t)-\vartheta _6 \vartheta _4(t)\right) [\left(\vartheta _1^2+\vartheta _3^2\right) b(t)\nonumber\\&+&\left(\vartheta
   _1+\vartheta _3\right) \vartheta _4'(t)]]/[\left(\vartheta _1+\vartheta _3\right) \vartheta _6] \, dt,
\end{eqnarray}}with  $\vartheta _1 \left(\vartheta _1+\vartheta _3\right) \vartheta _6\neq 0$.
{\begin{eqnarray}(III): \vartheta _7&=&\frac{\vartheta _2 \vartheta _6}{\vartheta _1}, \vartheta_8=-\frac{\vartheta _1 \vartheta _3}{\vartheta _6},
\vartheta _{5}(t)=\vartheta _{5},\nonumber\\
\vartheta_4(t)&=& \eta _5+\int_1^t [\left(\vartheta _1+\vartheta
_2-\vartheta _3\right) \left(\vartheta _3^2-\vartheta _6^2\right)
\vartheta _1^2 b(t)]/[\vartheta _1^2 \vartheta _3^2+\left(\vartheta
_1+\vartheta _2\right){}^2
   \vartheta _6^2] \, dt, \nonumber\\ \vartheta_{9}(t)&=&\eta _6+\int_1^t [\vartheta _1 \left(\vartheta _3^2-\vartheta _6^2\right) [\vartheta _3 \vartheta _1^2+\left(\vartheta _1+\vartheta _2\right) \vartheta
   _6^2] b(t)]/[\left(\vartheta
   _1+\vartheta _2\right){}^2 \vartheta _6^3+\vartheta _1^2 \vartheta _3^2 \vartheta _6] \, dt,\nonumber\\a(t)&=&-\frac{\vartheta _1 \left(2 \vartheta _1^2+2 \vartheta _2 \vartheta _1+\vartheta _2^2\right) \vartheta _3^2 \vartheta _{5} b(t)}{3 \vartheta _2 \left(\vartheta
   _1^2+\vartheta _6^2\right) \left(\vartheta _1^2 \vartheta _3^2+\left(\vartheta _1+\vartheta _2\right){}^2 \vartheta _6^2\right)}
\end{eqnarray}}with  $\vartheta _1 [\left(\vartheta
   _1+\vartheta _2\right){}^2 \vartheta _6^3+\vartheta _1^2 \vartheta _3^2 \vartheta _6]\neq 0$.
   {\begin{eqnarray}(IV): \vartheta _7&=&-\frac{\vartheta _1 \vartheta _2}{\vartheta _6},
   \vartheta_8=-\frac{\vartheta _1 \vartheta _3}{\vartheta _6}, \vartheta _{5}(t)=0,\nonumber\\
\vartheta_4(t)&=& \eta _7+\int_1^t \frac{\left(\vartheta
_1-\vartheta _2-\vartheta _3\right) \left(\vartheta _3^2-\vartheta
_6^2\right) b(t)}{\left(\vartheta _2+\vartheta
_3\right){}^2+\vartheta _6^2} \, dt,\nonumber\\
\vartheta_{9}(t)&=&\eta _8+\int_1^t \frac{\left(\vartheta
_3^2-\vartheta _6^2\right) [\vartheta _6^2+\vartheta _1
\left(\vartheta _2+\vartheta _3\right)] b(t)}{\vartheta
   _6^3+\left(\vartheta _2+\vartheta _3\right){}^2 \vartheta _6} \, dt
\end{eqnarray}}with  $\vartheta
   _6^3+\left(\vartheta _2+\vartheta _3\right){}^2 \vartheta _6\neq 0$.
{\begin{eqnarray}(V): \vartheta _7&=&-\frac{\vartheta _1 \vartheta _2}{\vartheta _6}, \vartheta_8=\frac{\vartheta _3 \vartheta _6}{\vartheta _1},
 \vartheta _{5}(t)=0,\nonumber\\
\vartheta_4(t)&=& \eta _9-\int_1^t [\vartheta _6^2 [\left(\vartheta
_1-\vartheta _2+\vartheta _3\right) \left(\vartheta _1^2+\vartheta
_3^2\right) b(t)]]/[\vartheta _1^2 \vartheta _2^2+\left(\vartheta
_1+\vartheta
   _3\right){}^2 \vartheta _6^2] \, dt,\nonumber\\ \vartheta_{9}(t)&=&\eta _{10}-\int_1^t \frac{\left(\vartheta _1^2+\vartheta _3^2\right) \vartheta _6 \left(\vartheta _2 \vartheta _1^2+\left(\vartheta _1+\vartheta _3\right) \vartheta
   _6^2\right) b(t)}{\vartheta _2^2 \vartheta _1^3+\left(\vartheta _1+\vartheta _3\right){}^2 \vartheta _6^2 \vartheta _1} \, dt
\end{eqnarray}}with  $\vartheta _2^2 \vartheta _1^3+\left(\vartheta _1+\vartheta _3\right){}^2 \vartheta _6^2 \vartheta _1\neq 0$.
{\begin{eqnarray}(VI): \vartheta _8&=&-\frac{\vartheta _2 \left(\vartheta _1+\vartheta _2+\vartheta _3\right)}{\vartheta _7}-\vartheta _6-\vartheta _7,
 \vartheta _{5}(t)=\vartheta _{5},\nonumber\\
\vartheta_4(t)&=& \eta _{11}+\int_1^t [\vartheta _7^4+2 \vartheta _6
\vartheta _7^3-\left(\vartheta _1^2-2 \vartheta _2 \vartheta _1-2
\vartheta _2^2+\vartheta _3^2-2 \vartheta
   _6^2\right) \vartheta _7^2\nonumber\\&+&2 \vartheta _2 \left(2 \vartheta _1+\vartheta _2\right) \vartheta _6 \vartheta _7+\vartheta _2^2 \left(\vartheta _1+\vartheta
   _2-\vartheta _3\right) \left(\vartheta _1+\vartheta _2+\vartheta _3\right)] b(t)\nonumber\\&/&[\left(\vartheta _1+\vartheta _2+\vartheta _3\right) \left(\vartheta _2^2+\vartheta _7^2\right)] \, dt,\nonumber\\ \vartheta_{9}(t)&=&\eta _{12}+\int_1^t [\left(\vartheta _2+2 \vartheta _3\right) \vartheta _7^4+2 \left(-\vartheta _1+\vartheta _2+\vartheta _3\right) \vartheta _6 \vartheta
   _7^3+\vartheta _2 [-\vartheta _1^2\nonumber\\&+&2 \left(\vartheta _2+\vartheta _3\right) \vartheta _1+2 \vartheta _2^2+\vartheta _3^2+2 \vartheta _6^2+4 \vartheta _2
   \vartheta _3] \vartheta _7^2+2 \vartheta _2^2 (\vartheta _1+\vartheta _2\nonumber\\&+&\vartheta _3) \vartheta _6 \vartheta _7+\vartheta _2^3 \left(\vartheta
   _1+\vartheta _2+\vartheta _3\right){}^2] b(t)/[\left(\vartheta _1+\vartheta _2+\vartheta _3\right) \vartheta _7 (\vartheta _2^2+\vartheta _7^2)] \, dt,\nonumber\\ a(t)&=&-[[\vartheta _7^4+2 \vartheta _6 \vartheta _7^3+2 [\vartheta _6^2+\vartheta _2 \left(\vartheta _1+\vartheta _2\right)] \vartheta _7^2+2
   \vartheta _2 \left(2 \vartheta _1+\vartheta _2\right) \vartheta _6 \vartheta _7\nonumber\\&+&\vartheta _2^2 \left(2 \vartheta _1^2+2 \vartheta _2 \vartheta _1+\vartheta
   _2^2\right)] \vartheta _{5} b(t)]\nonumber\\&/&[3 \left(\vartheta _1^2+\vartheta _6^2\right) \left(\vartheta _1 \vartheta _2+\vartheta _6 \vartheta _7\right)
   \left(\vartheta _2^2+\vartheta _7^2\right)],
\end{eqnarray}}with  $\vartheta _7\neq 0$.
{\begin{eqnarray}(VII): \vartheta _7&=&-\frac{\vartheta _3 \left(\vartheta _1+\vartheta _2+\vartheta _3\right)}{\vartheta _8}-\vartheta _6-\vartheta _8,
\vartheta _{5}(t)=\vartheta _{5},\nonumber\\
\vartheta_4(t)&=& \eta _{13}+\int_1^t [\vartheta _8^4 b(t)-\vartheta
_8^2 [\left(\vartheta _1^2-\vartheta _3^2-\vartheta _6^2\right)
b(t)]+2 \vartheta _1 \vartheta _3 \vartheta _6 \vartheta _8
b(t)]\nonumber\\&/&[\left(\vartheta _1+\vartheta _2+\vartheta
_3\right) \left(\vartheta _3^2+\vartheta _8^2\right)] \,
dt,\nonumber\\ \vartheta_{9}(t)&=&\eta _{14}-\int_1^t
\frac{\vartheta _8 \left(\vartheta _3 \vartheta _1^2+2 \vartheta _6
\vartheta _8 \vartheta _1+\vartheta _3 \left(\vartheta
_3^2-\vartheta
   _6^2+\vartheta _8^2\right)\right) b}{\left(\vartheta _1+\vartheta _2+\vartheta _3\right) \left(\vartheta _3^2+\vartheta _8^2\right)}\, dt,\nonumber\\ a(t)&=&[\vartheta _8 [\vartheta _3^4+\vartheta _1^2 \vartheta _3^2+2 \vartheta _1 \vartheta _6 \vartheta _8 \vartheta _3+\vartheta _8^4+\left(2 \vartheta
   _3^2+\vartheta _6^2\right) \vartheta _8^2] b(t) \vartheta _5]\nonumber\\&/&[3 \left(\vartheta _1^2+\vartheta _6^2\right) \left(\vartheta _3^2+\vartheta _8^2\right)
   [\vartheta _6 \vartheta _8^2+\left(\vartheta _6^2-\vartheta _1 \vartheta _2\right) \vartheta _8\nonumber\\&+&\vartheta _3 \left(\vartheta _1+\vartheta _2+\vartheta
   _3\right) \vartheta _6]],
\end{eqnarray}}with  $\vartheta _8\neq 0$,  $\eta _i (i=1,\cdots,14)$ is integral constant.
Substituting Eq. (5), Eq. (6), Eq. (7), Eq. (8), Eq. (9), Eq. (10)
and Eq. (11)  into the transformation
$u=\frac{6}{\rho}\,[ln\xi(x,y,z,t)]_x$, respectively. Abundant lump
solutions for Eq. (1) can be derived.

\includegraphics[scale=0.4,bb=-20 270 10 10]{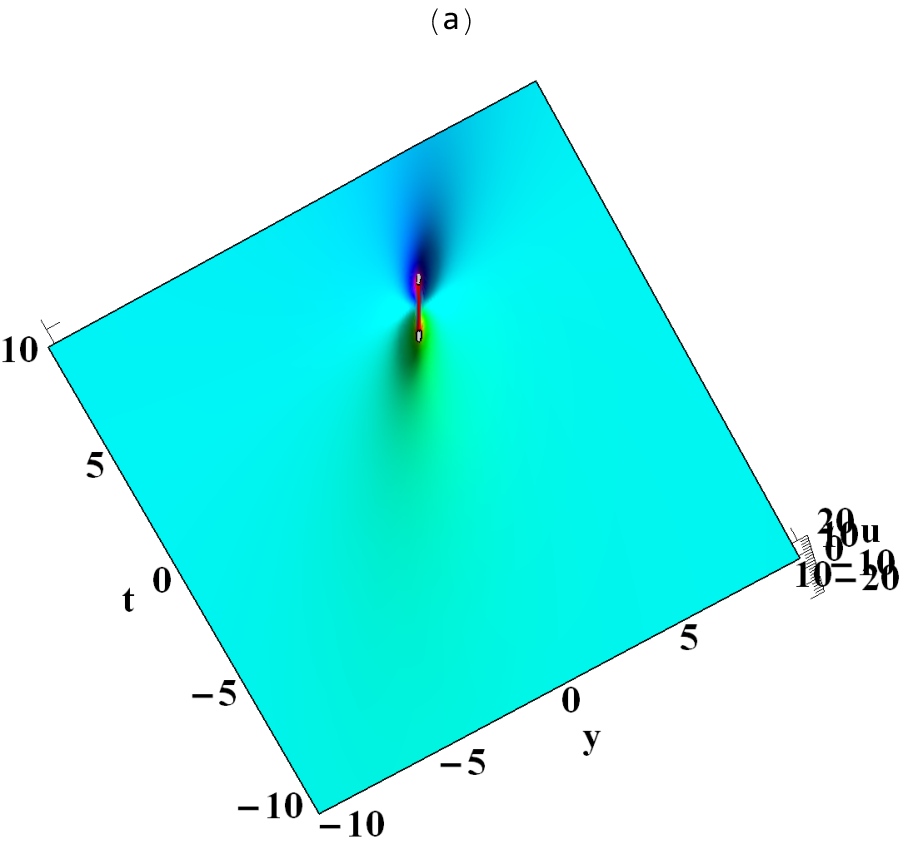}
\includegraphics[scale=0.4,bb=-355 270 10 10]{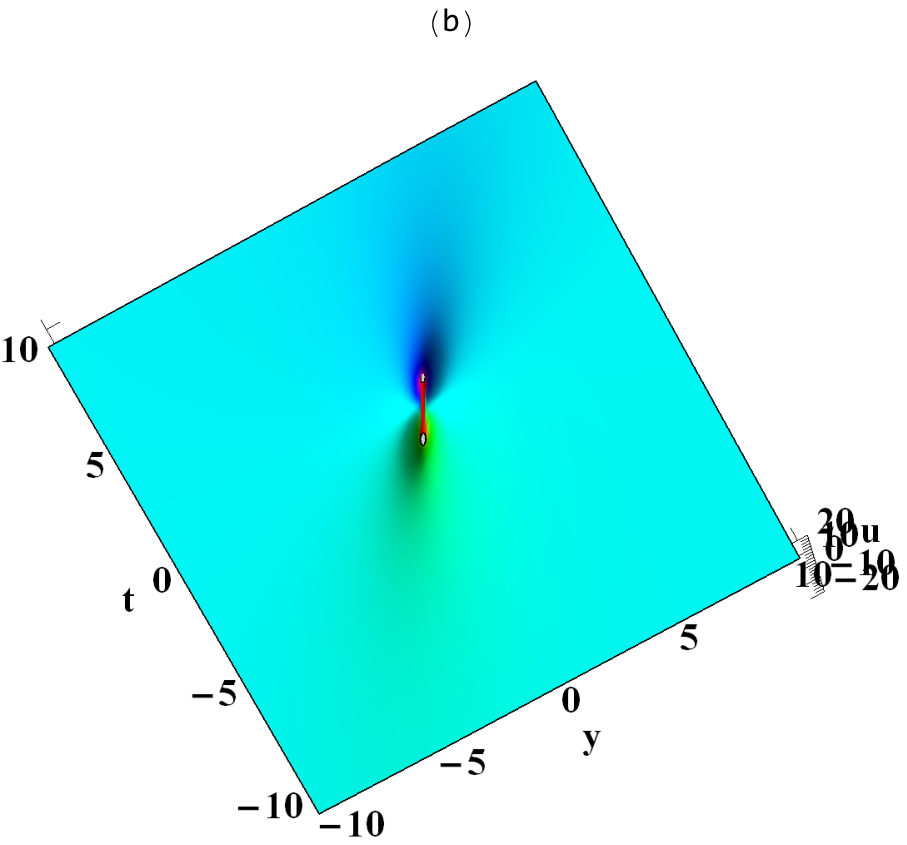}
\includegraphics[scale=0.4,bb=-360 270 10 10]{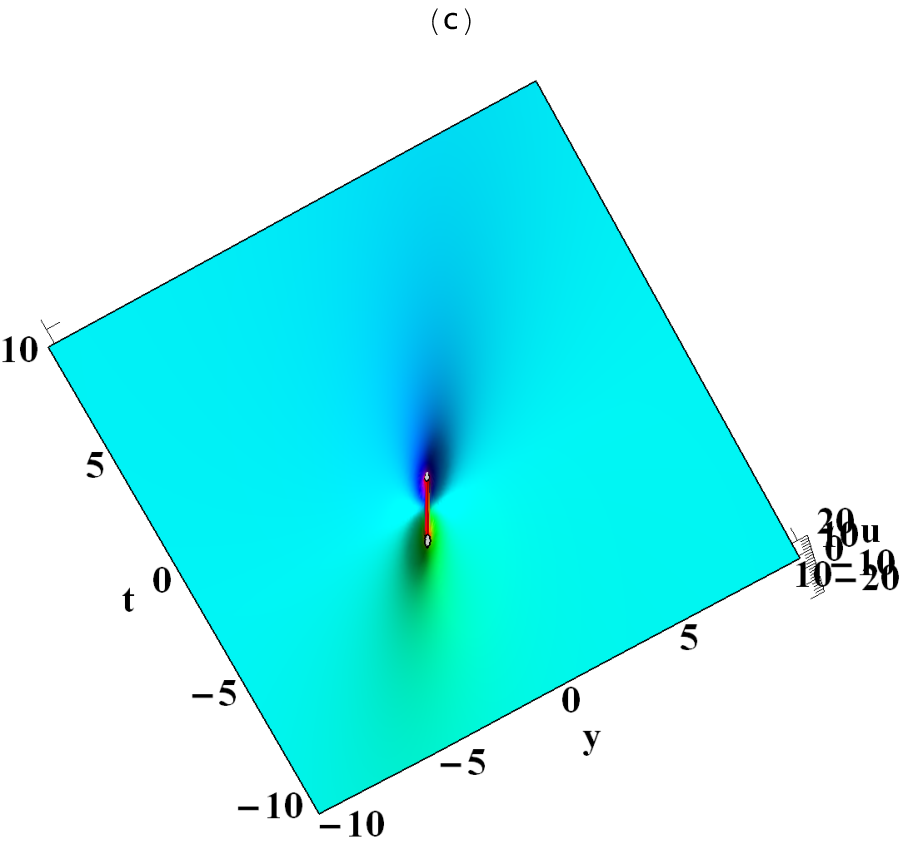}
\includegraphics[scale=0.35,bb=960 620 10 10]{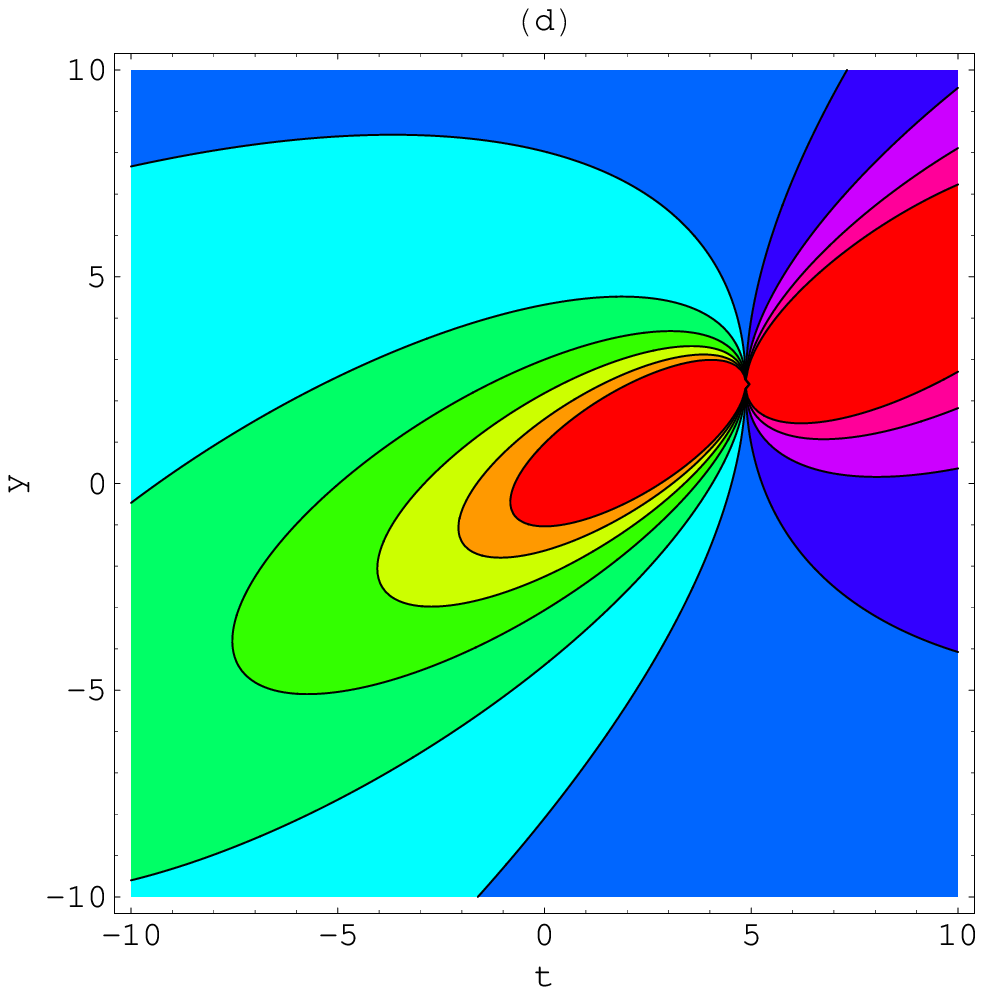}
\includegraphics[scale=0.35,bb=-405 620 10 10]{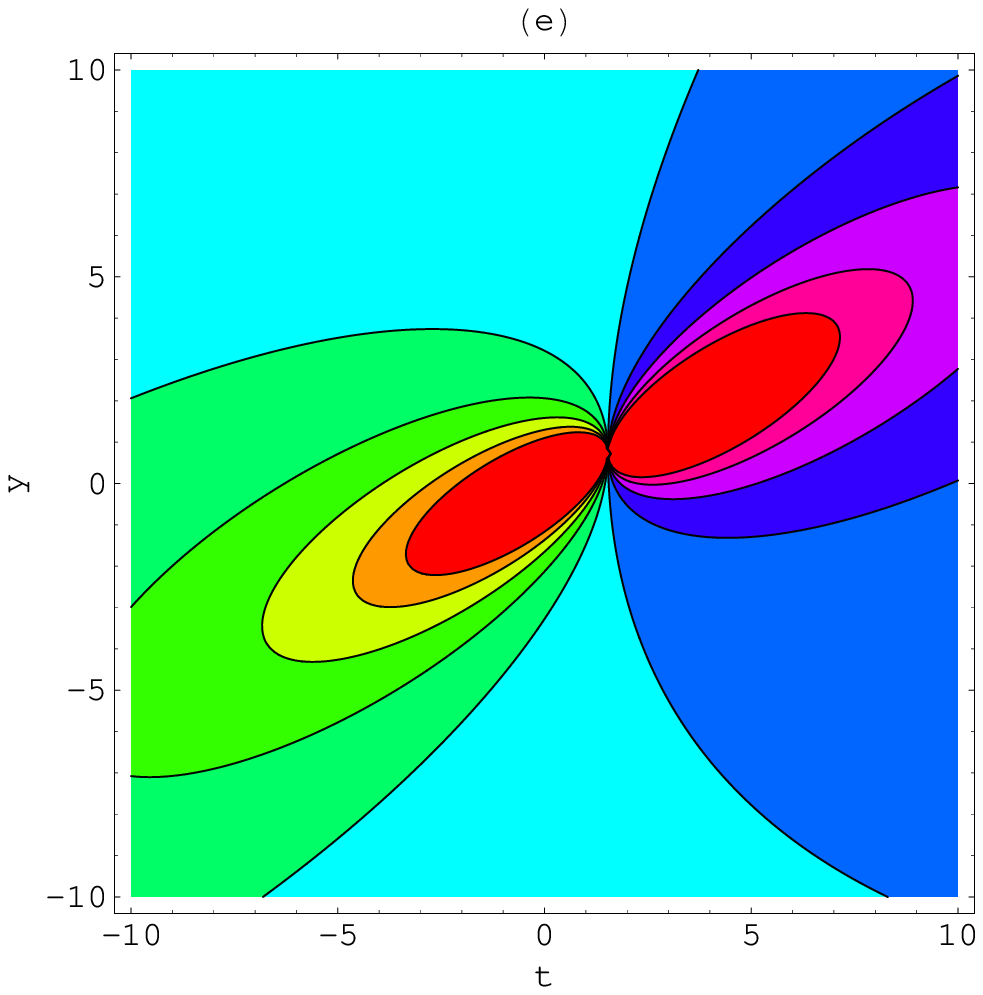}
\includegraphics[scale=0.35,bb=-400 620 10 10]{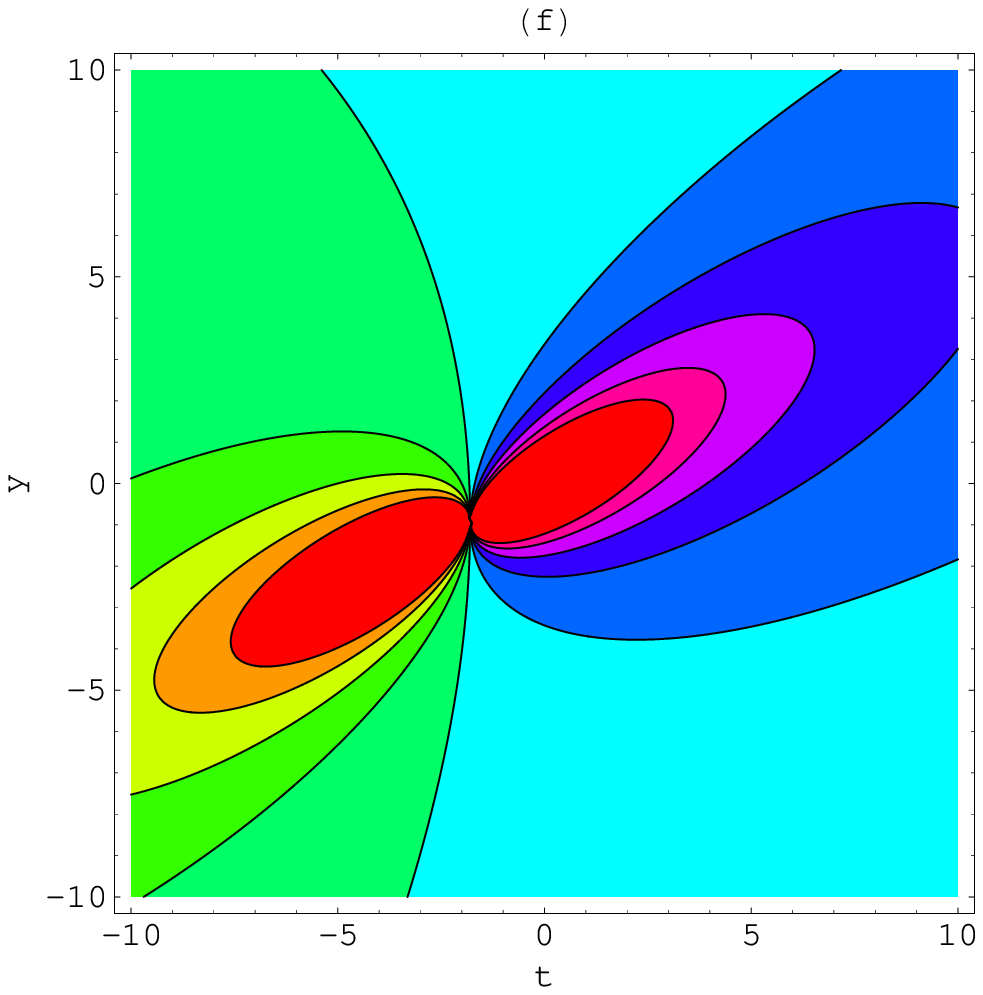}
\vspace{7.5cm}
\begin{tabbing}
\textbf{Fig.1}. Solution (12) in Eq. (13) with $b(t)=1$  when $x=
-5$  in (a) (d), $x= 0$ in (b) (e)\\ and $x = 5$ in (c) (f).\\
\end{tabbing}

\quad As a example, substituting Eq. (8) into
$u=\frac{6}{\rho}\,[ln\xi(x,y,z,t)]_x$, we have {\begin{eqnarray}
u^{(IV)}&=&[6 [2 \vartheta _1 [\vartheta _4(t)+x \vartheta _1+y
\vartheta _2+z \vartheta _3]+2 \vartheta _6 (\vartheta
   _{9}(t)+x \vartheta _6\nonumber\\&-&\frac{y \vartheta _1 \vartheta _2}{\vartheta _6}-\frac{z \vartheta _1 \vartheta _3}{\vartheta _6})]]
   [\rho  [[
\vartheta _4(t)+x \vartheta _1+y \vartheta _2+z \vartheta
_3]{}^2\nonumber\\&+&[\vartheta _{9}(t)+x \vartheta _6-\frac{y
\vartheta _1
   \vartheta _2}{\vartheta _6}-\frac{z \vartheta _1 \vartheta _3}{\vartheta _6}]{}^2]].
\end{eqnarray}}

\quad In order to analysis the dynamical behaviors for solution
(12), suppose that {\begin{eqnarray} \vartheta_1=3, \vartheta_2=-3,
\vartheta_3=2, \rho=1, \vartheta_6=-1, \eta_7=\eta_8=z=0.
\end{eqnarray}}Substituting Eq. (13) into Eq. (12), the dynamical behaviors for solution (12) are shown in Fig. 1 and Fig. 2.

\includegraphics[scale=0.4,bb=-20 270 10 10]{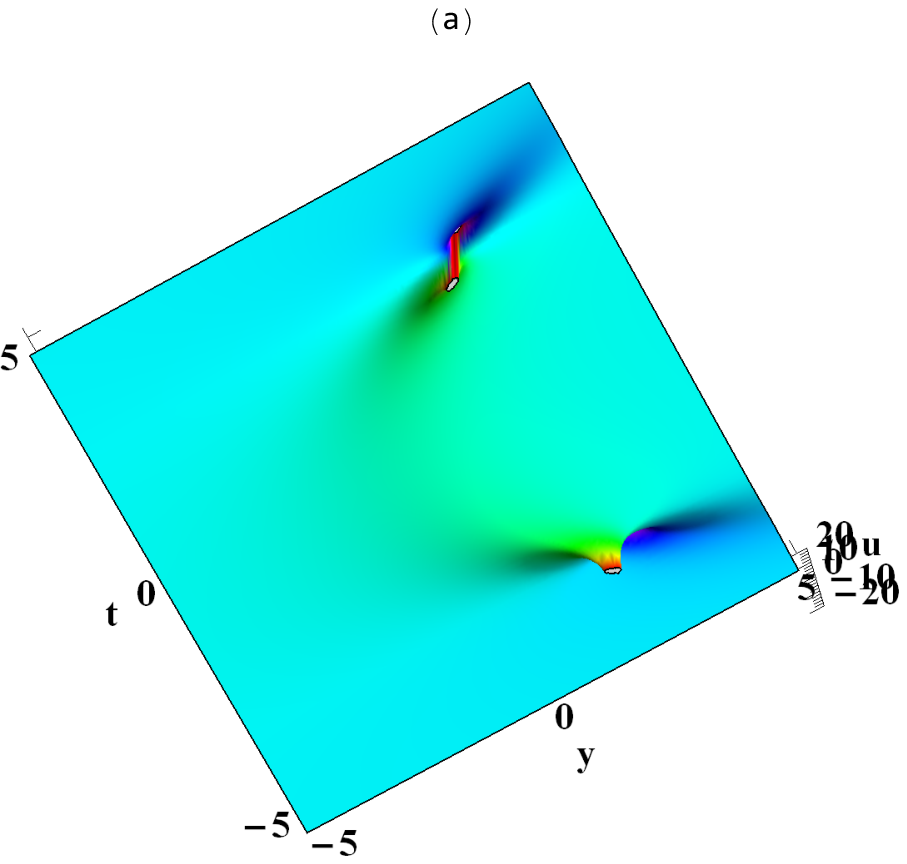}
\includegraphics[scale=0.4,bb=-355 270 10 10]{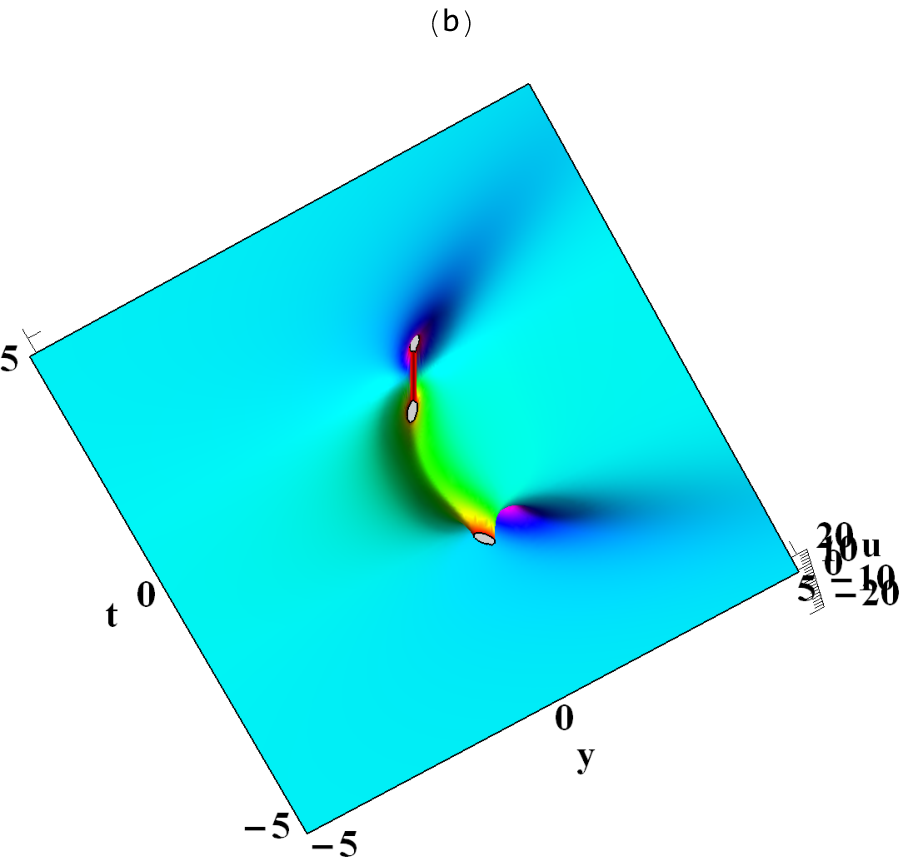}
\includegraphics[scale=0.4,bb=-360 270 10 10]{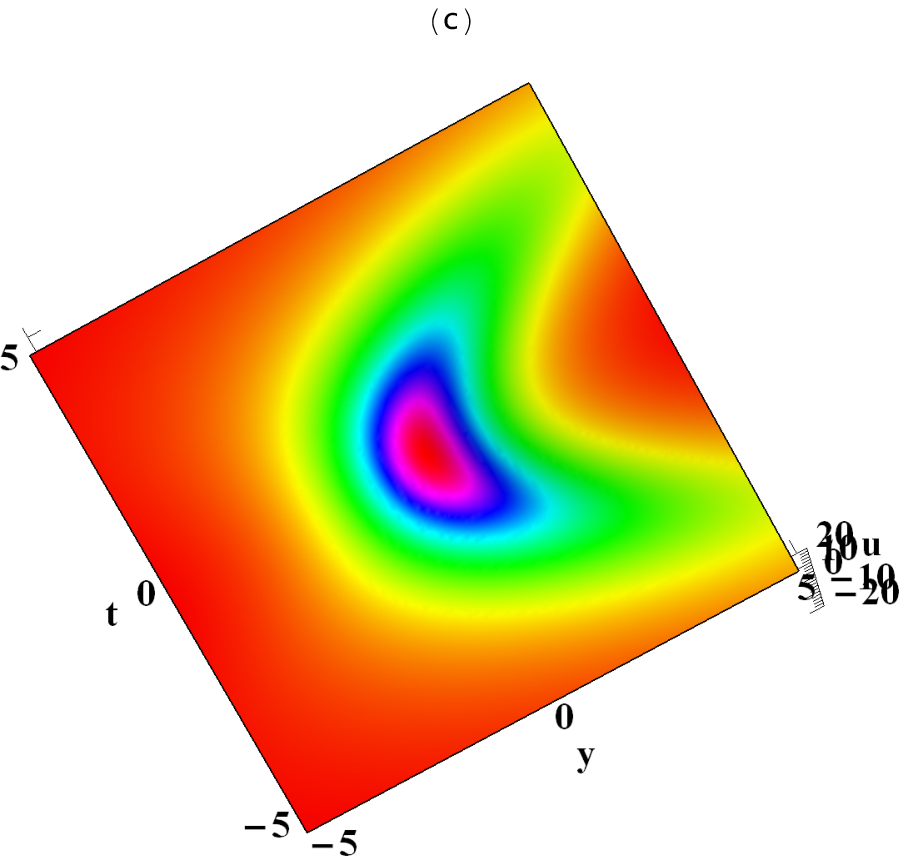}
\includegraphics[scale=0.35,bb=960 620 10 10]{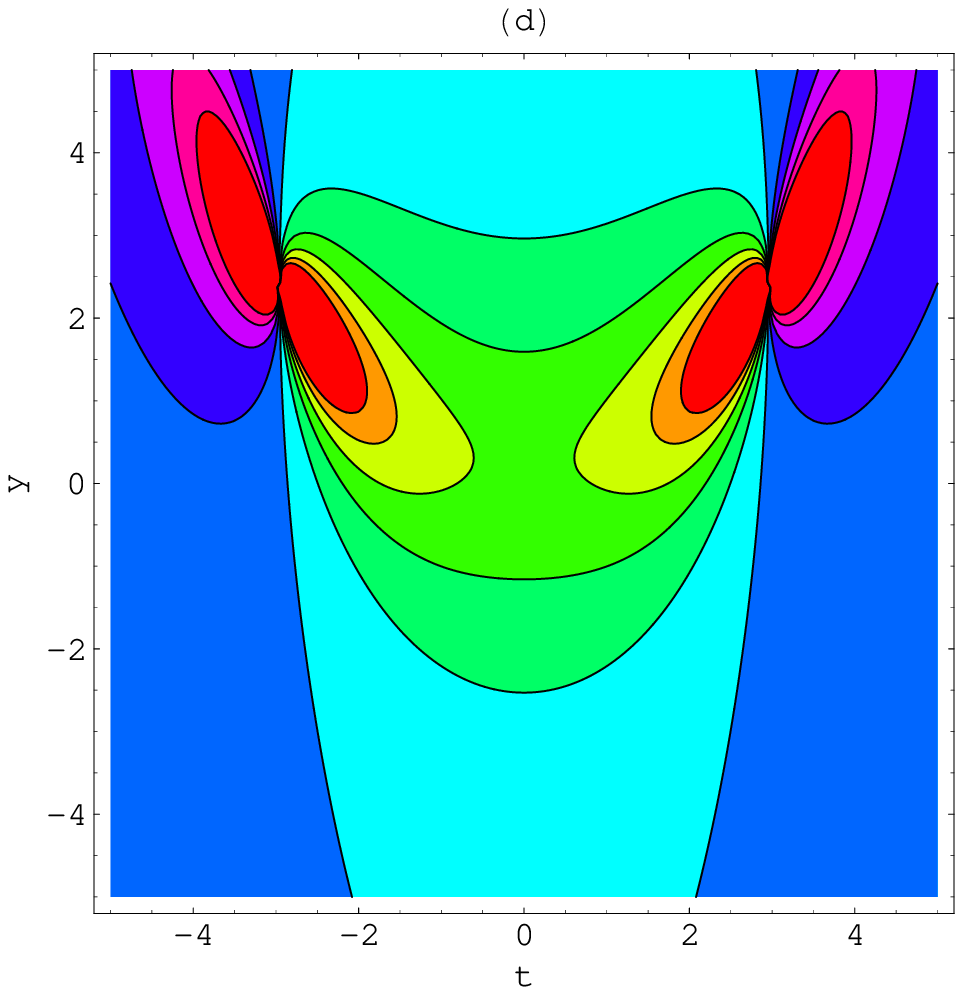}
\includegraphics[scale=0.35,bb=-405 620 10 10]{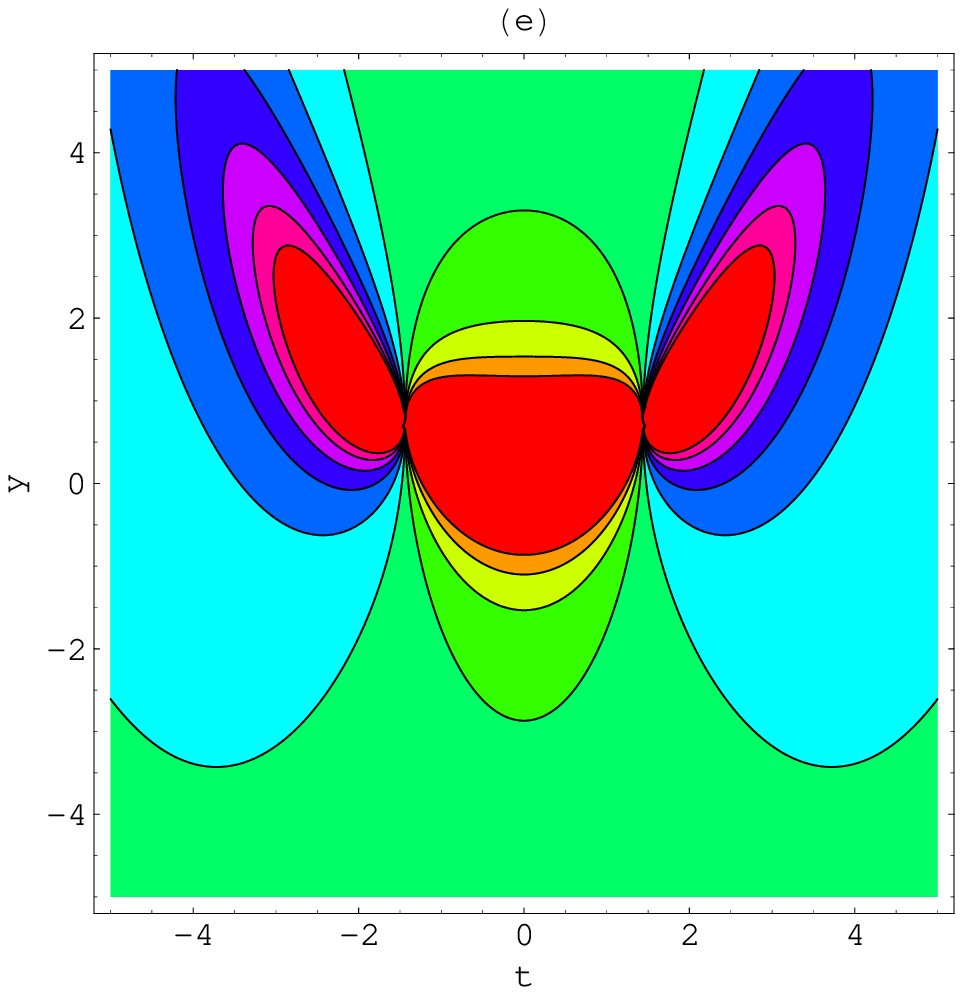}
\includegraphics[scale=0.35,bb=-400 620 10 10]{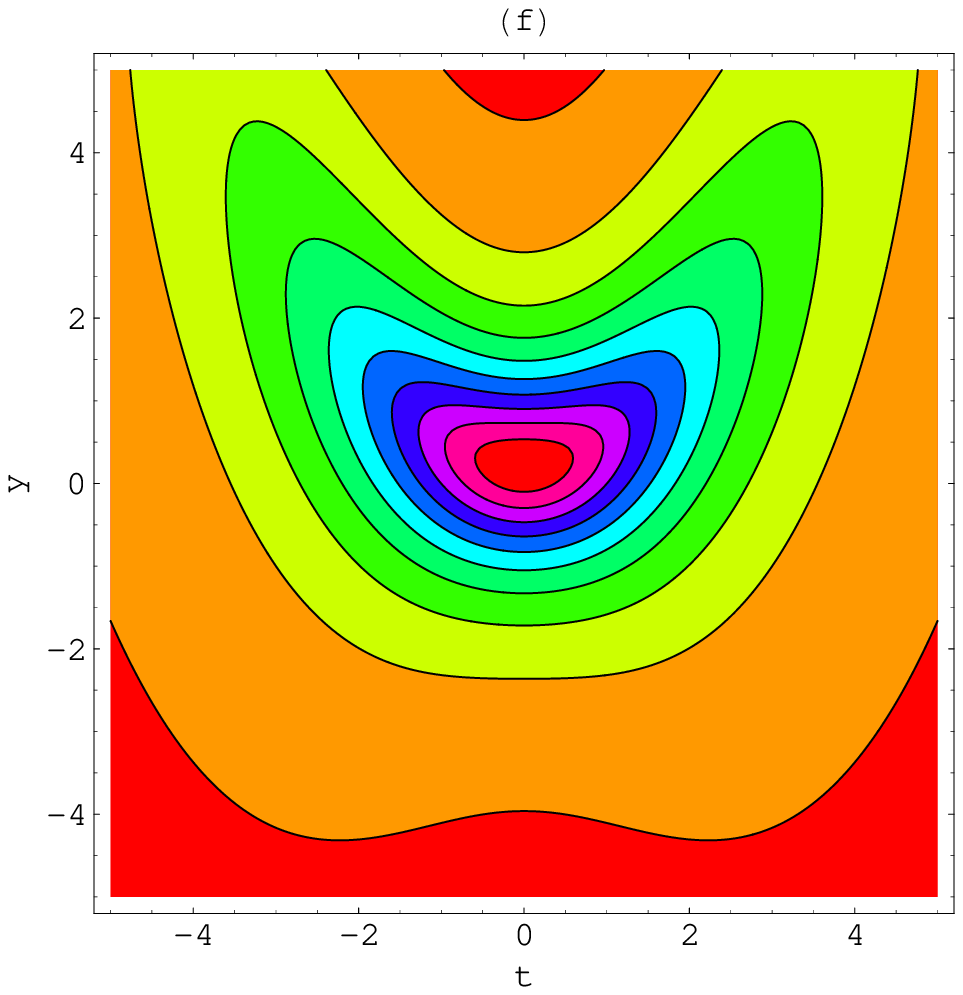}
\vspace{7.5cm}
\begin{tabbing}
\textbf{Fig.2}. Solution (12) in Eq. (13) with $b(t)=t$  when $x=
-5$  in (a) (d), $x= 0$ in (b) (e)\\ and $x = 5$ in (c) (f).\\
\end{tabbing}

\quad In Fig. 1, lump consists of a small hole and one high peak can
be observed. When $x$ sets different values, the lump's amplitude
and velocity keep unchange during the propagation. Fig. 2 shows the
interaction between two lump waves. When $x=-5$, two lump waves
propagate towards each other, their amplitudes do not
 alter before the interaction. When $x=0$, two lump waves collide together to form interactions, their amplitudes
become smaller. When $x=5$, two lump waves completely converge
together and propagate along the same direction.\\

\noindent {\bf 3. Lump-stripe solution}\\

\quad In order to research the interaction solution between the lump
solution and one stripe soliton (called Lump-stripe solution), a
mixed function with rational function and exponential function is
supposed  as follows {\begin{eqnarray} \zeta&=&\vartheta _4(t)+x \vartheta _1+y \vartheta _2+z \vartheta _3,\nonumber\\
\varsigma&=&\vartheta _{9}(t)+x \vartheta _6+y \vartheta _7+z \vartheta _8,\nonumber\\
\xi&=& \zeta^2+\varsigma^2+\vartheta _{5}(t)+\alpha _{14}(t)
e^{\vartheta _{13}(t)+\vartheta _{10} x+\vartheta _{11} y+\vartheta
_{12} z}.
\end{eqnarray}}Substituting Eq. (14) into Eq. (3) via the Mathematical
software, we have {\begin{eqnarray} \vartheta _7&=&-\frac{\vartheta
_1 \vartheta _2}{\vartheta _6},
   \vartheta_8=-\frac{\vartheta _1 \vartheta _3}{\vartheta _6}, \vartheta _{5}(t)=0, a(t)=\frac{2 \left(\vartheta _{11}^2+2 \vartheta _{12}
   \vartheta _{11}+2 \vartheta _{12}^2\right) b(t)}{\vartheta _{10} \vartheta _{11} [\vartheta _{10}^2-3
   \left(\vartheta _{11}+\vartheta _{12}\right){}^2]}, \nonumber\\
\vartheta_4(t)&=& \eta _{15}+\int_1^t \frac{\left(\vartheta
_1-\vartheta _2-\vartheta _3\right) \left(\vartheta _3^2-\vartheta
_6^2\right) b(t)}{\left(\vartheta _2+\vartheta
_3\right){}^2+\vartheta _6^2} \, dt,\nonumber\\
\vartheta_{9}(t)&=&\eta _{16}+\int_1^t \frac{\left(\vartheta
_3^2-\vartheta _6^2\right) [\vartheta _6^2+\vartheta _1
\left(\vartheta _2+\vartheta _3\right)] b(t)}{\vartheta
   _6^3+\left(\vartheta _2+\vartheta _3\right){}^2 \vartheta _6} \,
   dt,\nonumber\\ \vartheta_{13}(t)&=&\eta _{17}-\int_1^t \frac{\left(\vartheta _{10}-\vartheta _{11}-\vartheta _{12}\right)
    \left(\vartheta _{10}^2+3 \vartheta _{12}^2\right) b(t)}{\vartheta _{10}^2-3
   \left(\vartheta _{11}+\vartheta _{12}\right){}^2}+\frac{\alpha _{14}'(t)}{\alpha _{14}(t)} \,
   dt,\nonumber\\
   \vartheta_3&=&\frac{\vartheta _2 [\left(2 \vartheta _{11}^2+3 \vartheta _{12} \vartheta _{11}+
   2 \vartheta _{12}^2\right) \vartheta _{10}^2+3 \vartheta _{11} \vartheta
   _{12}^2 \left(\vartheta _{11}+\vartheta _{12}\right)]}{\vartheta _{11} [3 \vartheta _{11} \vartheta _{12} \left(\vartheta _{11}+\vartheta
   _{12}\right)-\vartheta _{10}^2 \left(\vartheta _{11}+2 \vartheta
   _{12}\right)]},\nonumber\\
   \vartheta_6&=&[\epsilon_1 [-[\vartheta _2^3 \vartheta _{10}^2 \left(\vartheta _{10}+\vartheta _{11}+\vartheta _{12}\right)
    \left(\vartheta _{11}^2+2 \vartheta _{12} \vartheta
   _{11}+2 \vartheta _{12}^2\right) [\left(\vartheta _{11}+2 \vartheta _{12}\right) \vartheta _{10}^2\nonumber\\&+&2 \vartheta _{11}
    \vartheta _{12} \vartheta _{10}+3
   \vartheta _{11} \vartheta _{12} \left(\vartheta _{11}+\vartheta _{12}\right)] [\left(5 \vartheta _{11}^2+4 \vartheta _{12}
   \vartheta _{11}+4 \vartheta
   _{12}^2\right) \vartheta _{10}^4\nonumber\\&-&6 \vartheta _{11} \left(\vartheta _{11}+2 \vartheta _{12}\right)
   \left(2 \vartheta _{11}^2+\vartheta _{12} \vartheta
   _{11}+\vartheta _{12}^2\right) \vartheta _{10}^2-27 \vartheta _{11}^2 \vartheta _{12}^2 \left(\vartheta _{11}+\vartheta _{12}\right){}^2]]
   \nonumber\\&/&[\vartheta
   _{11}^2 [3 \vartheta _{11} \vartheta _{12} \left(\vartheta _{11}+\vartheta _{12}\right)-\vartheta _{10}^2 \left(\vartheta _{11}+2 \vartheta
   _{12}\right)]{}^3]]^{0.5}]/[\sqrt{3} [[\vartheta _2 (\vartheta _{10}+\vartheta _{11}\nonumber\\&+&\vartheta _{12})
    \left(\vartheta _{11}^2+2 \vartheta
   _{12} \vartheta _{11}+2 \vartheta _{12}^2\right) [\left(\vartheta _{11}+2 \vartheta _{12}\right) \vartheta _{10}^2
   +2 \vartheta _{11} \vartheta _{12}
   \vartheta _{10}\nonumber\\&+&3 \vartheta _{11} \vartheta _{12} \left(\vartheta _{11}+\vartheta _{12}\right)]]/[3 \vartheta _{11} \vartheta _{12} \left(\vartheta
   _{11}+\vartheta _{12}\right)-\vartheta _{10}^2 \left(\vartheta _{11}+2 \vartheta
   _{12}\right)]]^{0.5}],
\end{eqnarray}}where $\eta _{15}$, $\eta _{16}$ and $\eta _{17}$ are integral constants, $\epsilon_1=\pm 1$. Substituting  Eq. (15) into the
     transformation $u=\frac{6}{\rho}\,[ln\xi(x,y,z,t)]_x$, the lump-stripe solution for Eq. (1) are obtained as follows
{\begin{eqnarray} u&=& [6 [\vartheta _{10} \alpha _{14}(t)
e^{\vartheta _{13}(t)+x \vartheta _{10}+y \vartheta _{11}+z
\vartheta _{12}}+2 \vartheta _1 [\vartheta _4(t)+x
   \vartheta _1+y \vartheta _2+z \vartheta _3]\nonumber\\&+&2 \vartheta _6 [\vartheta _9(t)+x \vartheta _6
   -\frac{y \vartheta _1 \vartheta _2}{\vartheta _6}-\frac{z
   \vartheta _1 \vartheta _3}{\vartheta _6}]]]/[\rho  [\alpha _{14}(t) e^{\vartheta _{13}(t)+x \vartheta _{10}+y \vartheta _{11}+z \vartheta
   _{12}}\nonumber\\&+&[\vartheta _4(t)+x \vartheta _1+y \vartheta _2+z \vartheta _3]{}^2+[\vartheta _9(t)+x \vartheta _6-\frac{y \vartheta _1 \vartheta
   _2}{\vartheta _6}-\frac{z \vartheta _1 \vartheta _3}{\vartheta
   _6}]{}^2]].
\end{eqnarray}}

\includegraphics[scale=0.45,bb=-20 230 10 10]{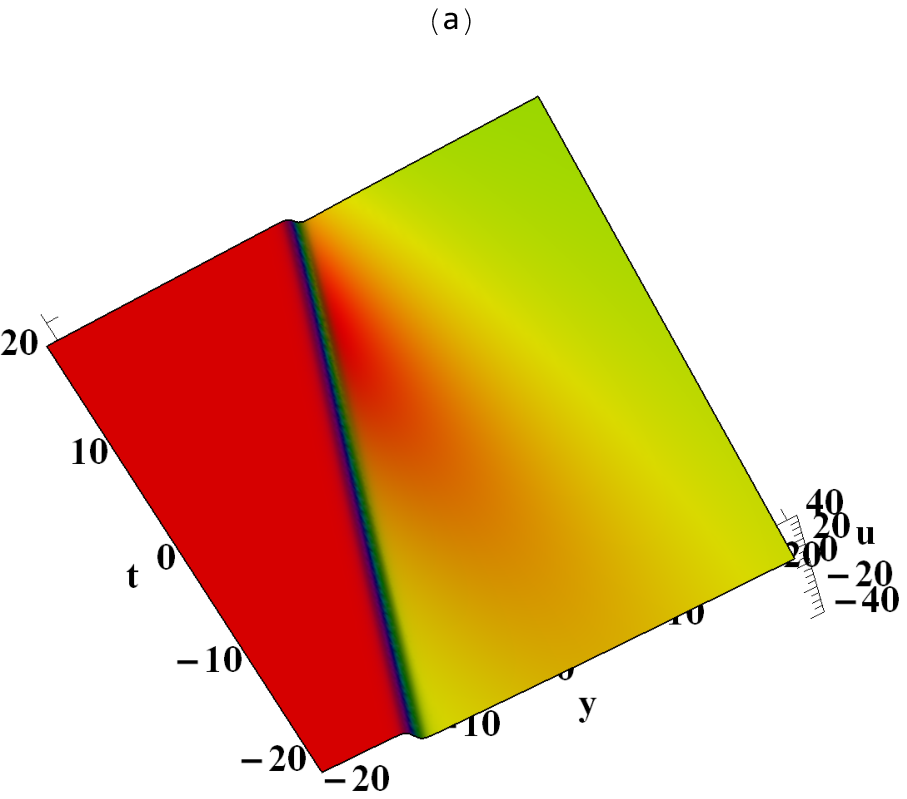}
\includegraphics[scale=0.45,bb=-305 230 10 10]{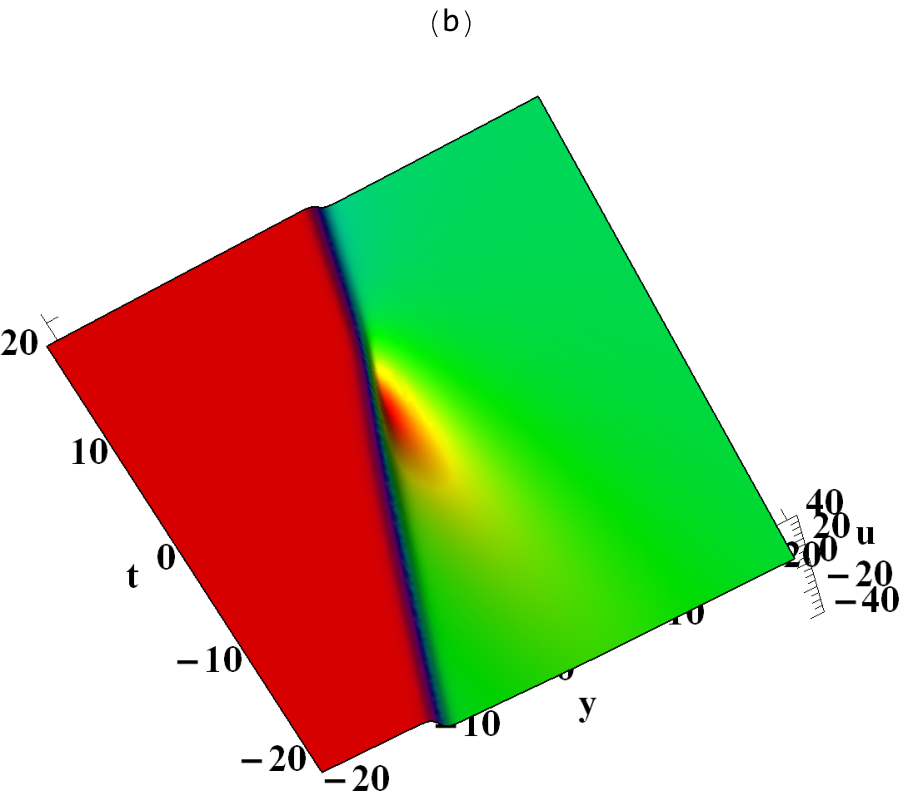}
\includegraphics[scale=0.45,bb=-300 230 10 10]{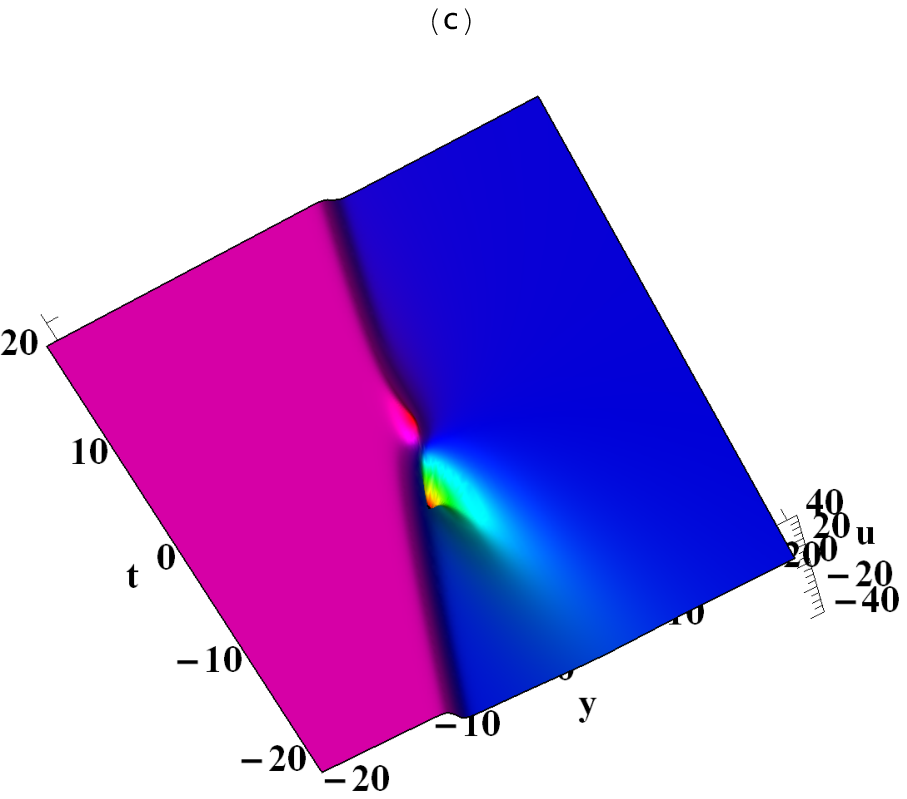}
\includegraphics[scale=0.45,bb=520 500 10 10]{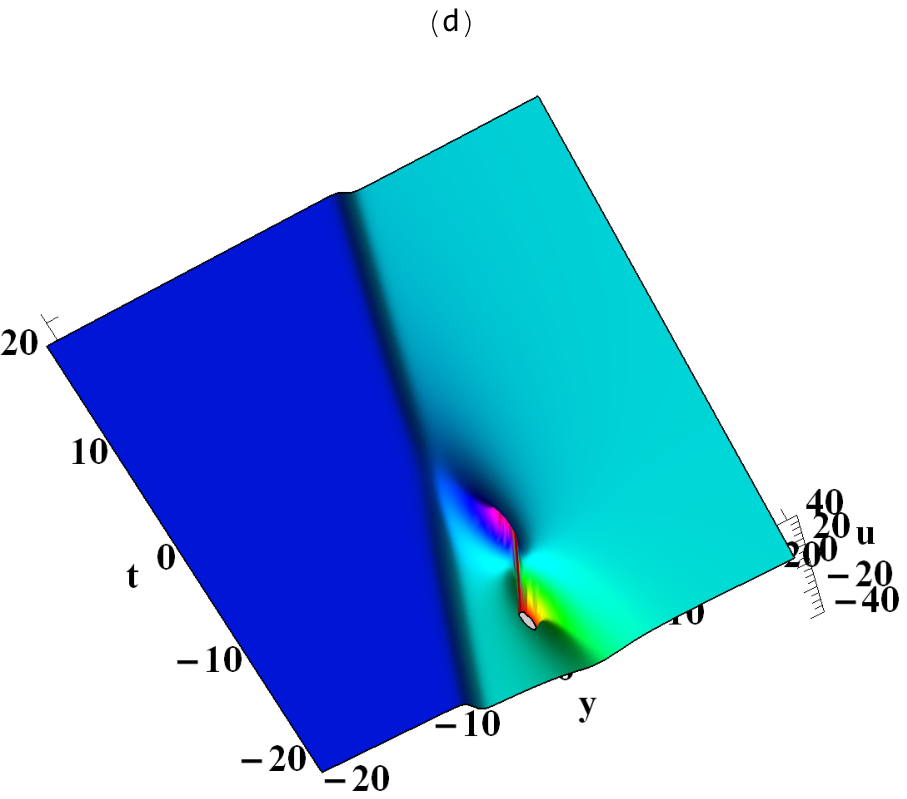}
\includegraphics[scale=0.45,bb=-400 500 10 10]{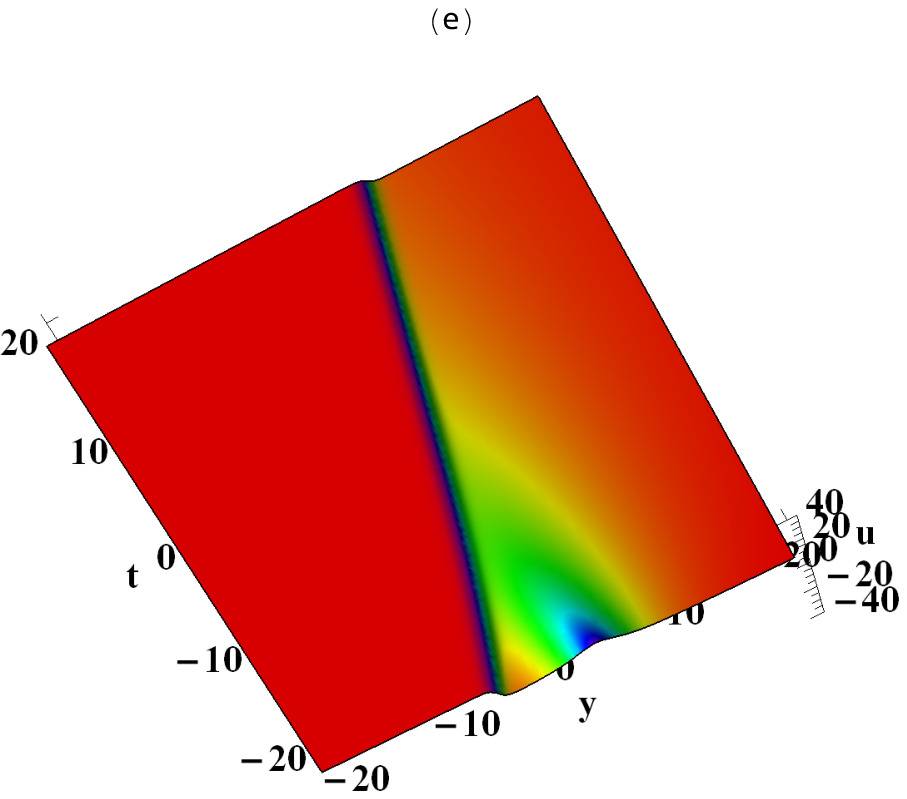}
\vspace{7.6cm}
\begin{tabbing}
\textbf{Fig.3}. Solution (16) in Eq. (17) with $b(t)=1$  when (a)
$x= -5$, (b) $x= -2$, (c) $x= 0$,\\ (d) $x= 2$, (e) $x= 5$.
\end{tabbing}

\includegraphics[scale=0.4,bb=80 270 10 10]{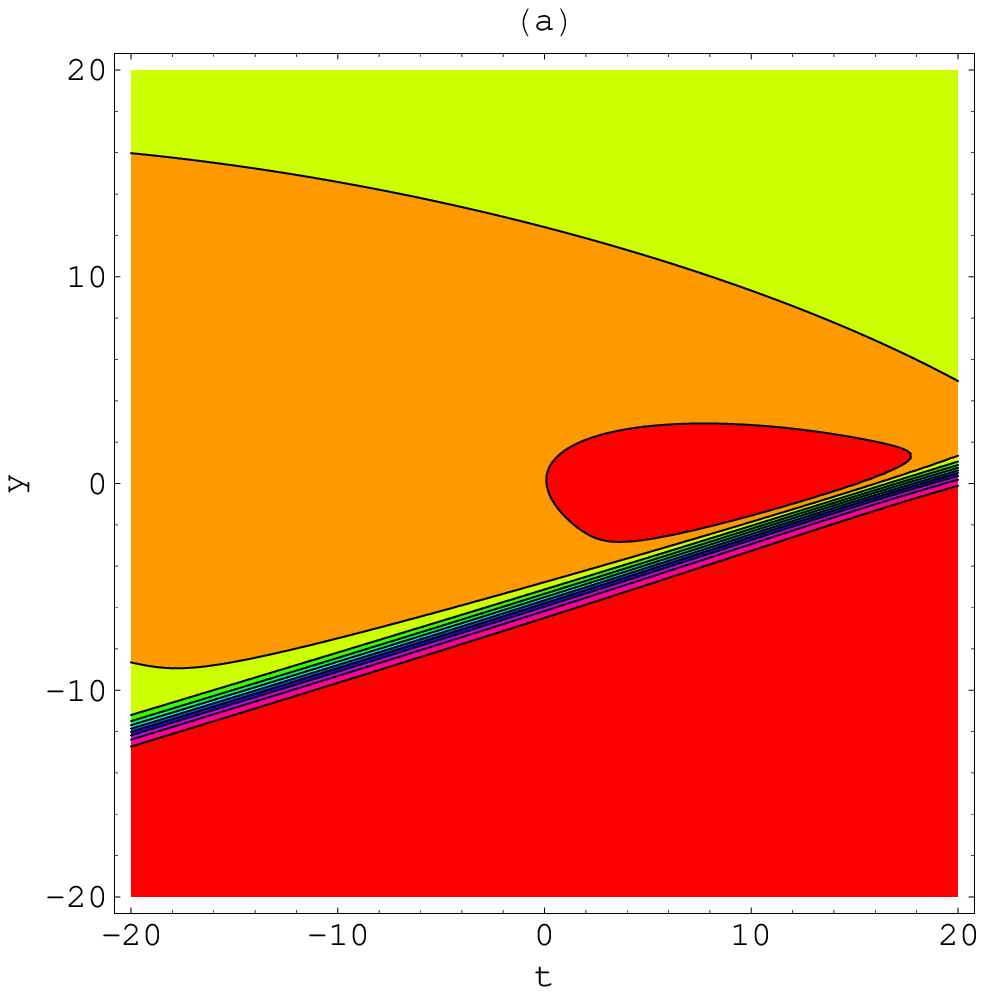}
\includegraphics[scale=0.4,bb=-335 270 10 10]{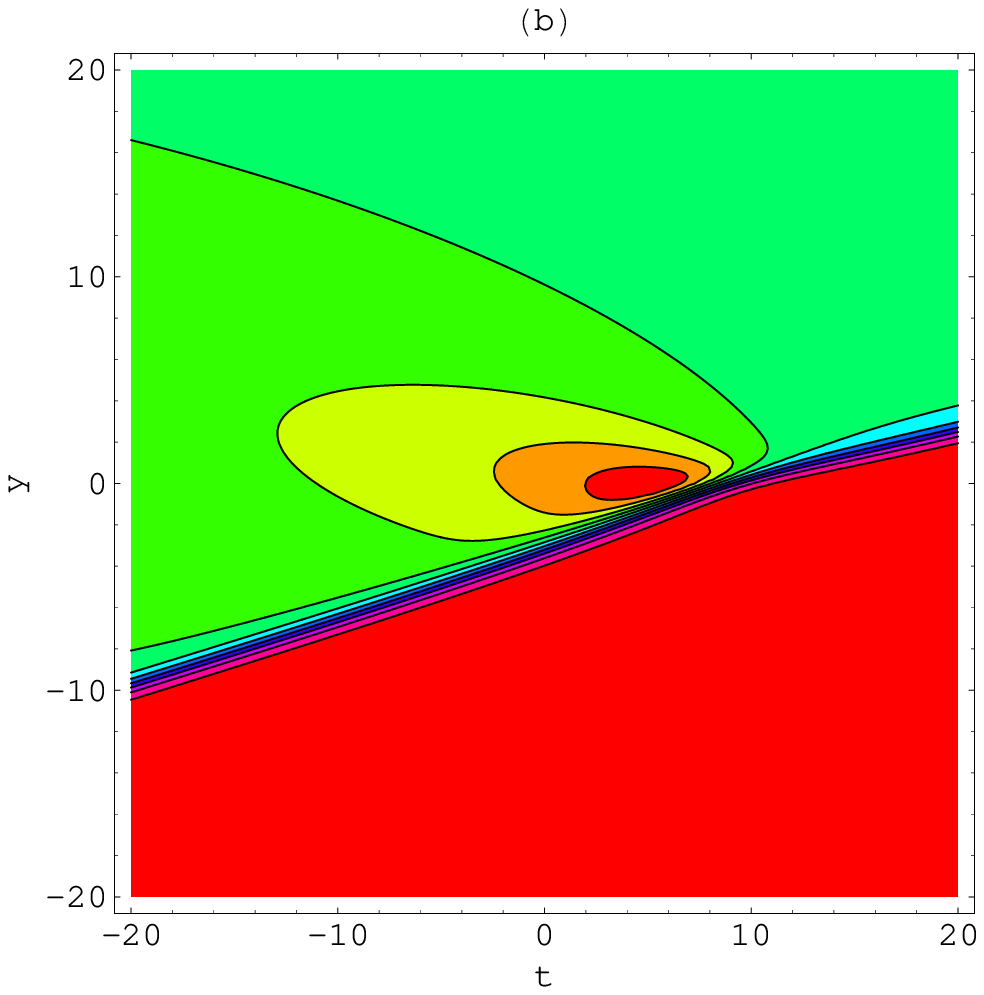}
\includegraphics[scale=0.4,bb=-340 270 10 10]{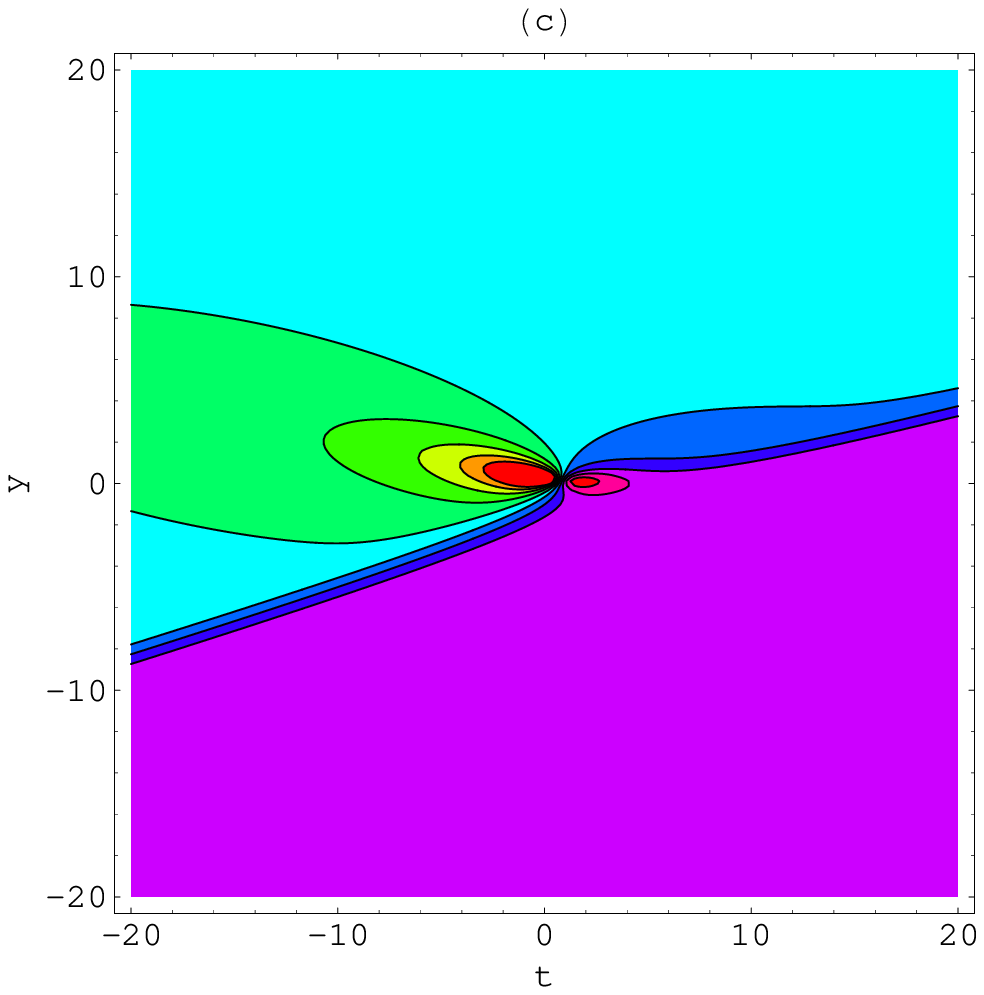}
\includegraphics[scale=0.4,bb=520 570 10 10]{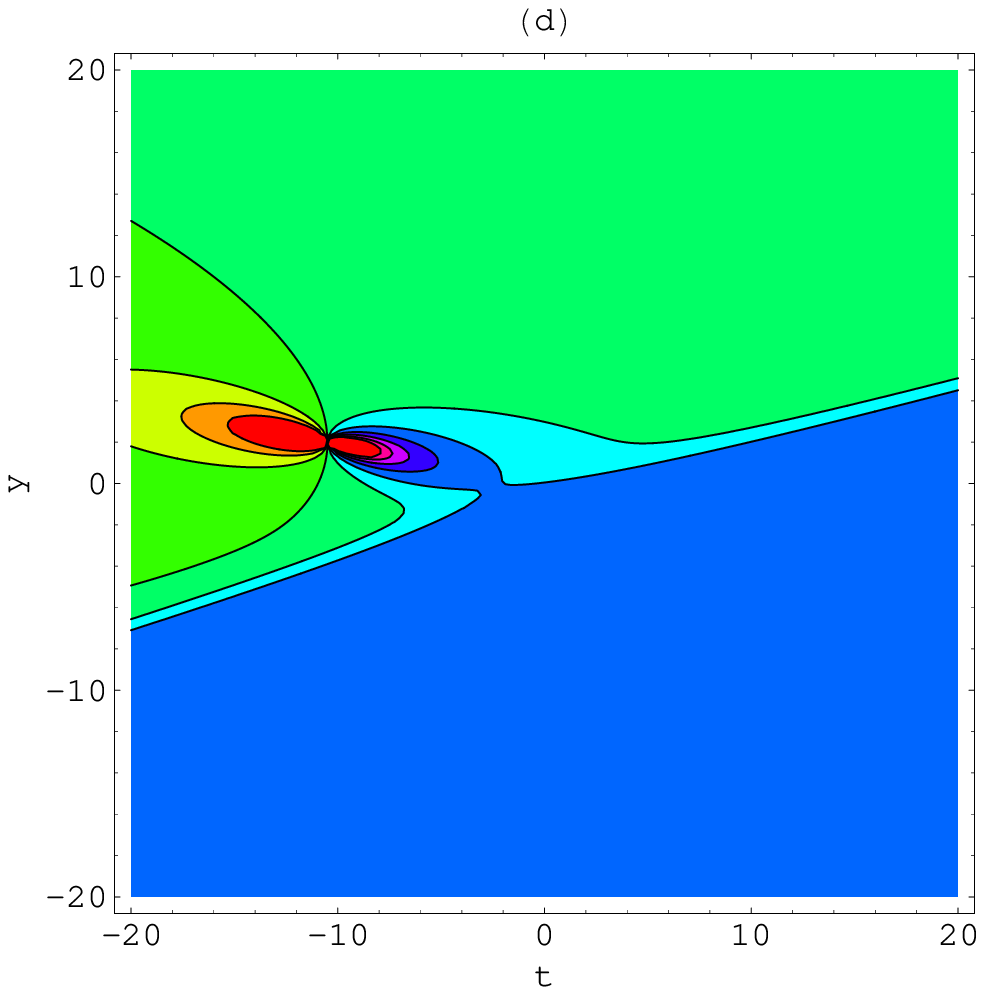}
\includegraphics[scale=0.4,bb=-400 570 10 10]{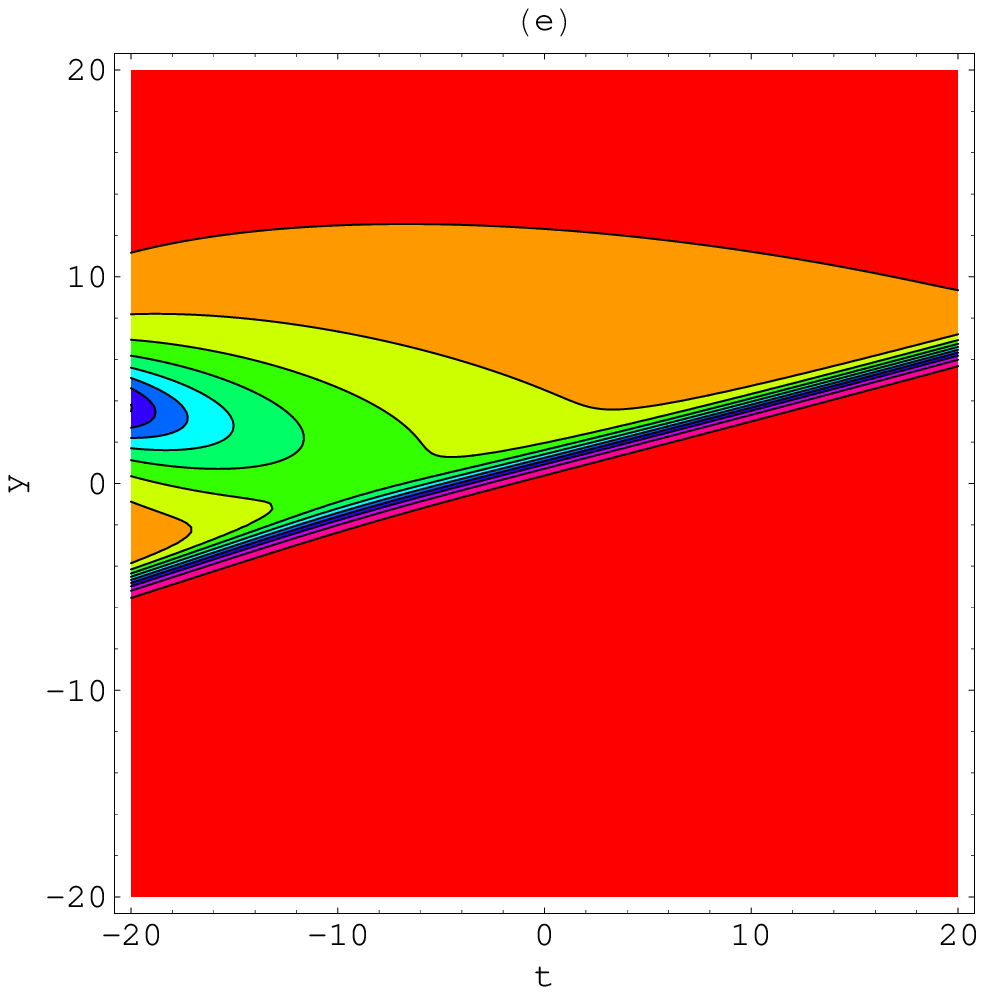}
\vspace{7.6cm}
\begin{tabbing}
\textbf{Fig.4}. The corresponding contour plots of Fig.3.\\
\end{tabbing}

 \quad Choosing the special values of the parameters as
{\begin{eqnarray} \vartheta_1&=&3, \vartheta_2=-1, \vartheta_{10}=2,
 \rho=\alpha_{14}(t)=\epsilon_1=1,\nonumber\\
\vartheta_{12}&=&\eta_{15}=\eta_{16}=\eta_{17}= z=0,
\vartheta_{11}=-3.
\end{eqnarray}}Substituting Eq. (17) into Eq. (16), the dynamical behaviors for solution (16) are shown in Fig. 3, Fig. 4 and Fig. 5.

\includegraphics[scale=0.4,bb=-20 270 10 10]{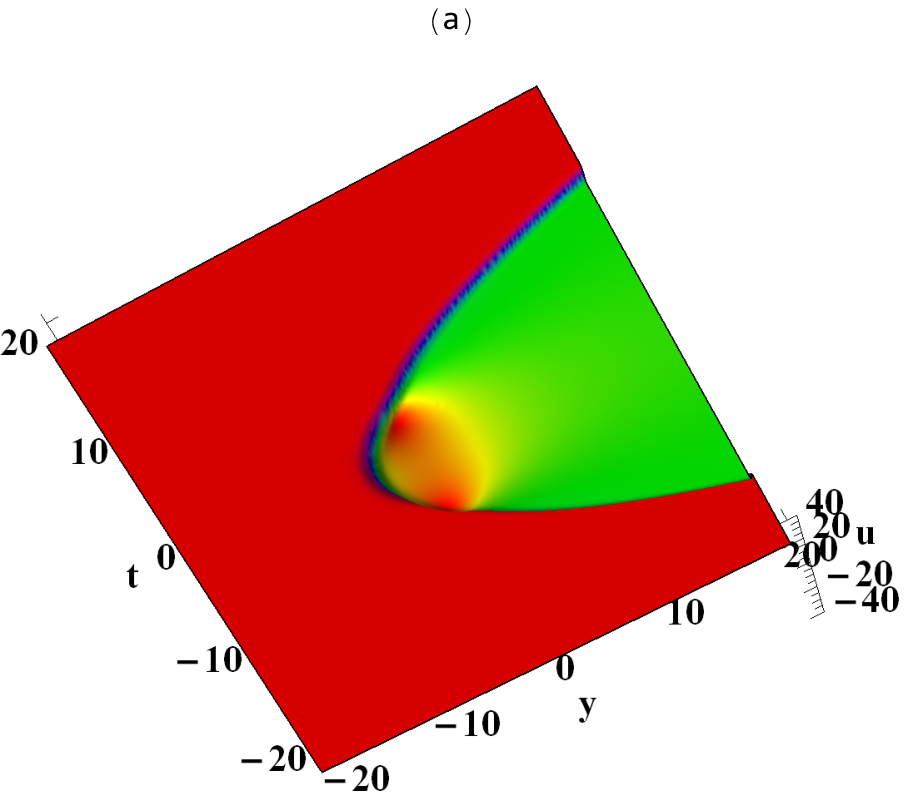}
\includegraphics[scale=0.4,bb=-355 270 10 10]{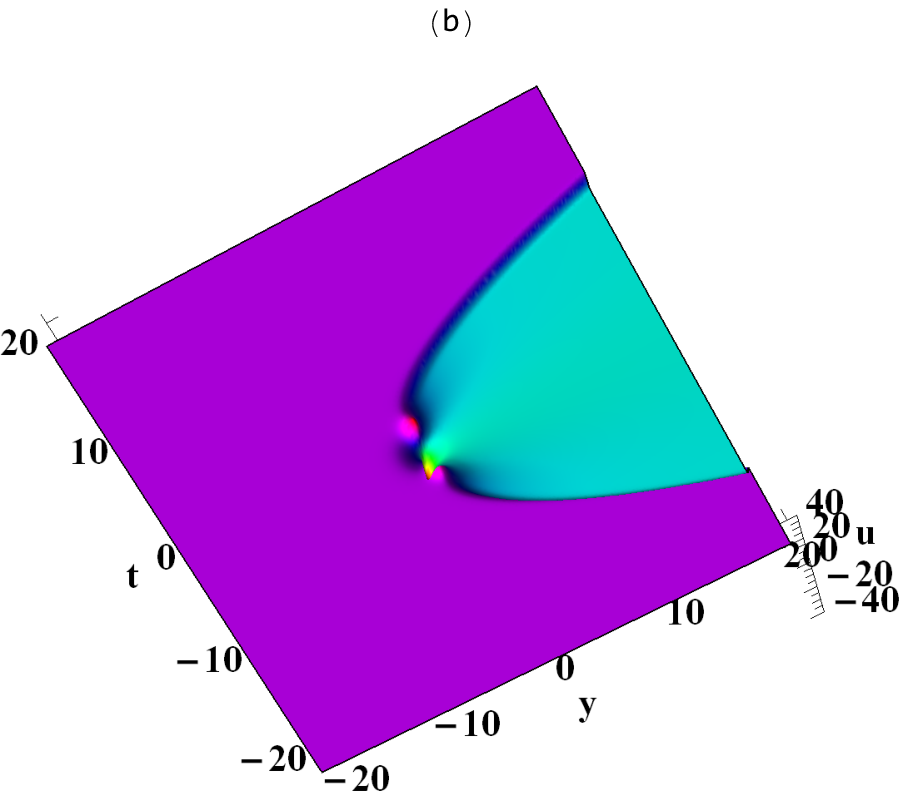}
\includegraphics[scale=0.4,bb=-360 270 10 10]{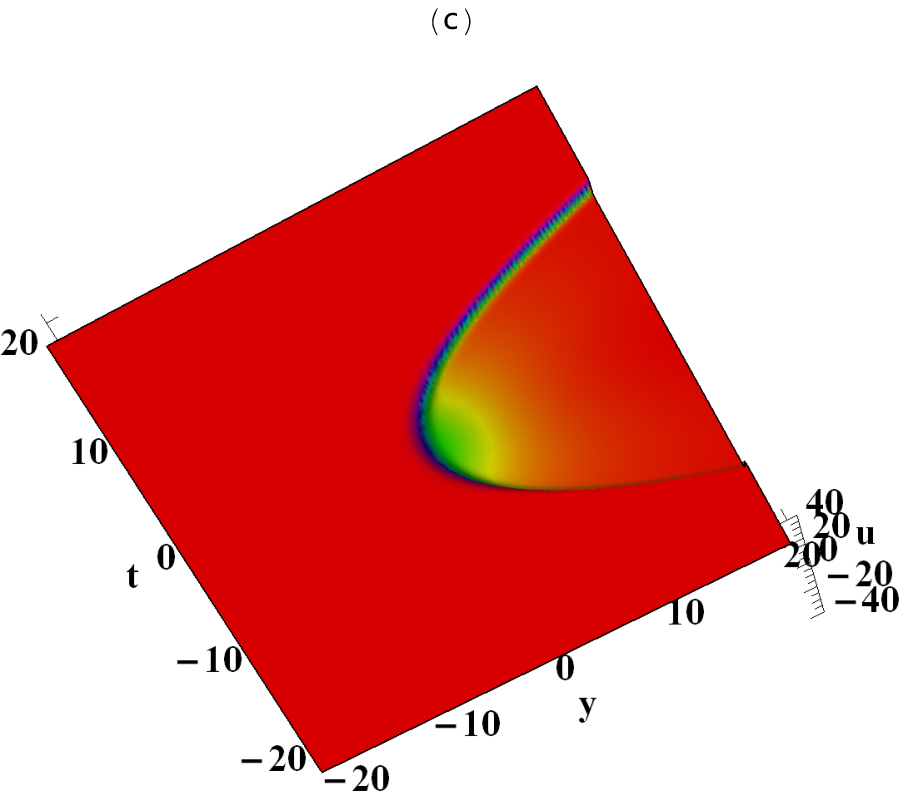}
\includegraphics[scale=0.35,bb=960 620 10 10]{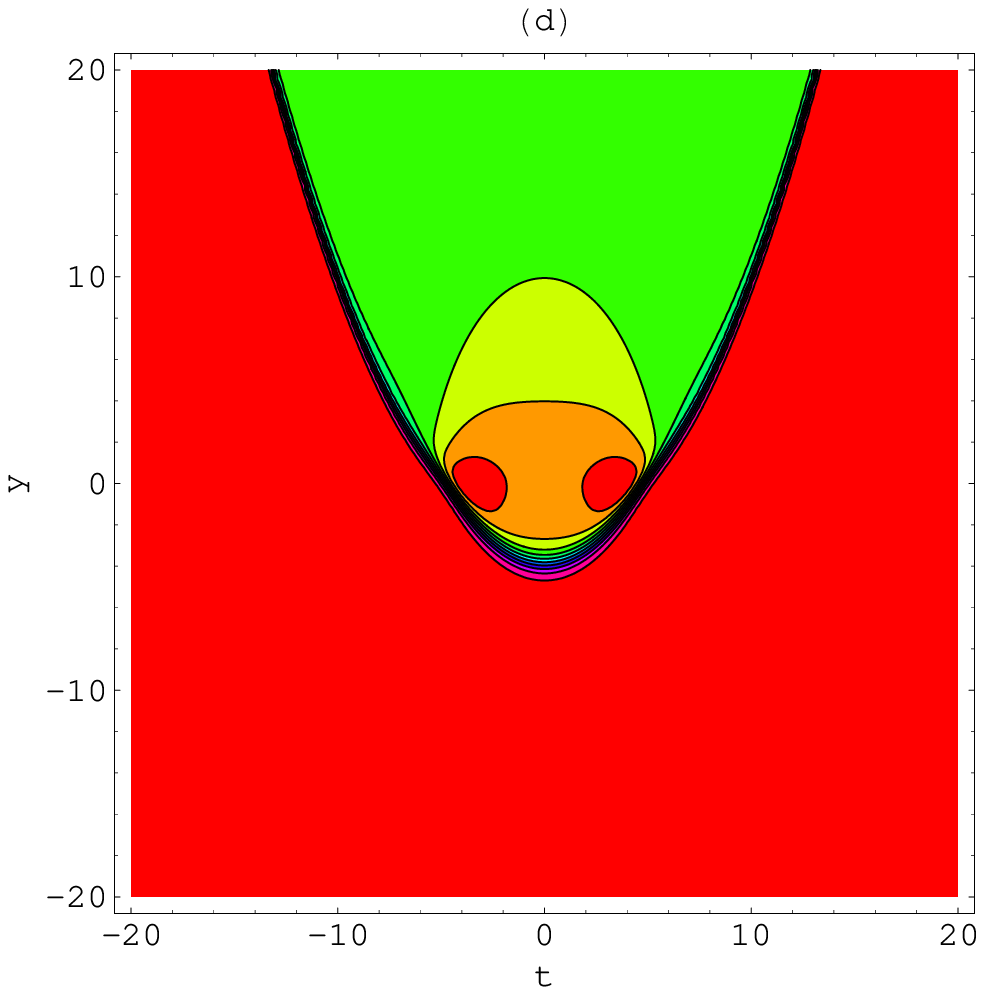}
\includegraphics[scale=0.35,bb=-405 620 10 10]{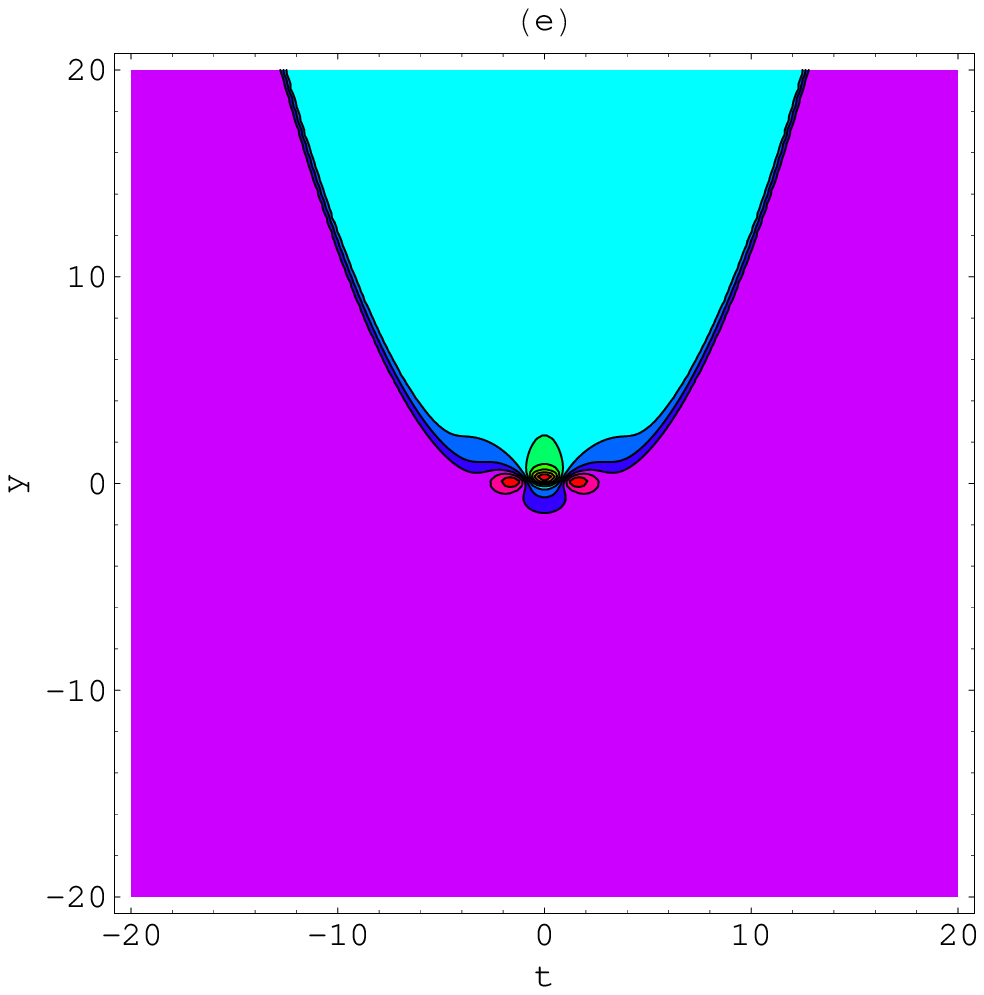}
\includegraphics[scale=0.35,bb=-400 620 10 10]{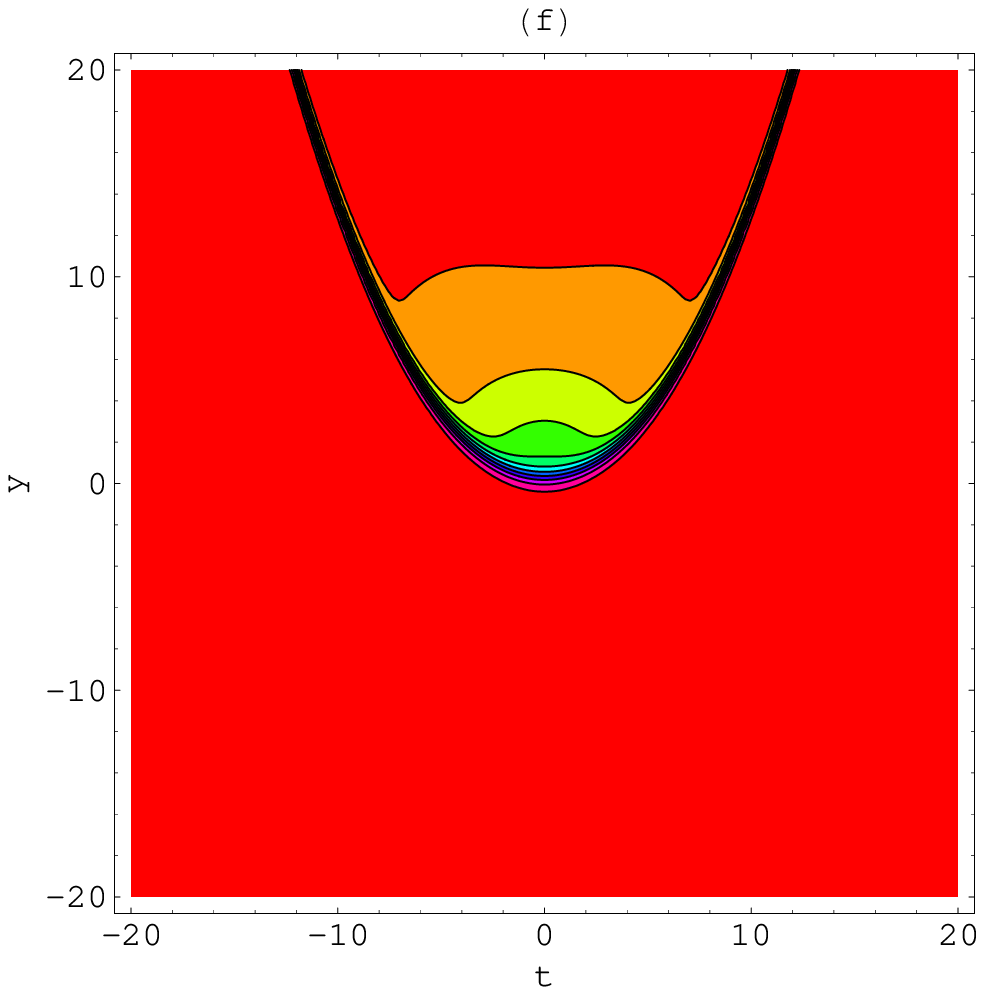}
\vspace{7.5cm}
\begin{tabbing}
\textbf{Fig.5}. Solution (16) in Eq. (17) with $b(t)=t$  when $x=
-3$  in (a) (d), $x= 0$ in (b) (e)\\ and $x = 3$ in (c) (f).\\
\end{tabbing}

\quad Fig. 3 describes the dynamical properties of the lump-stripe
solution with $b(t)=1$. Fig. 4 lists the corresponding contour plots
of Fig. 3. when $x=-5, -2, 0, 2, 5$, lump begins to cover the stripe
soliton  bit by bit, until they merge into together and go on
propagating. Fig. 5 shows the dynamical properties of solution (16)
with $b(t)=t$.

\includegraphics[scale=0.45,bb=-20 250 10 10]{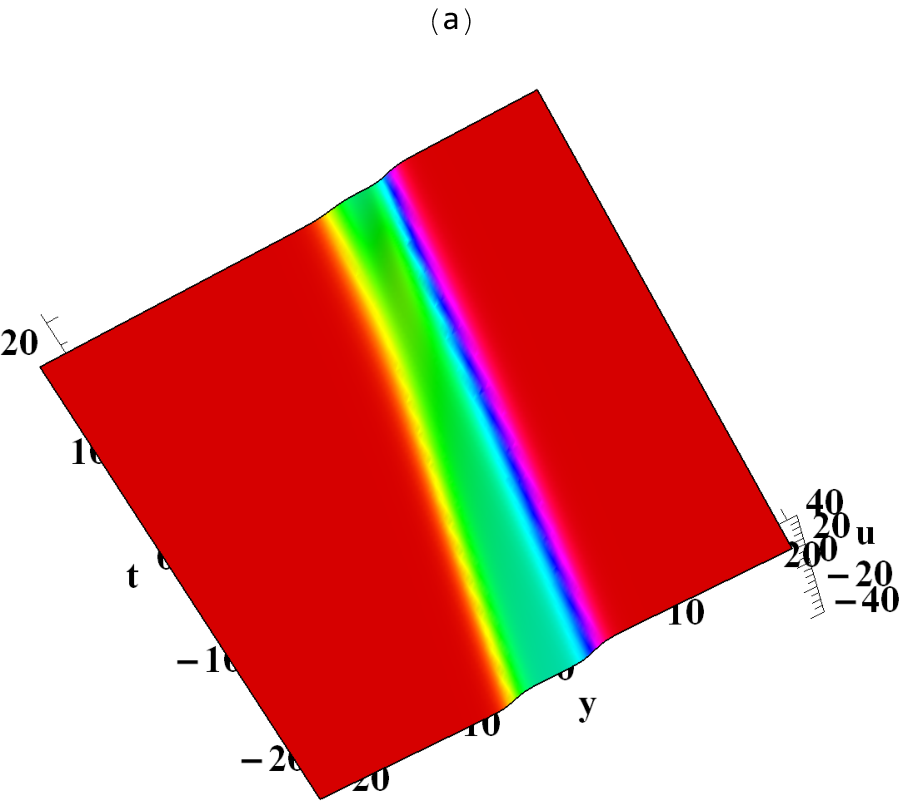}
\includegraphics[scale=0.45,bb=-305 250 10 10]{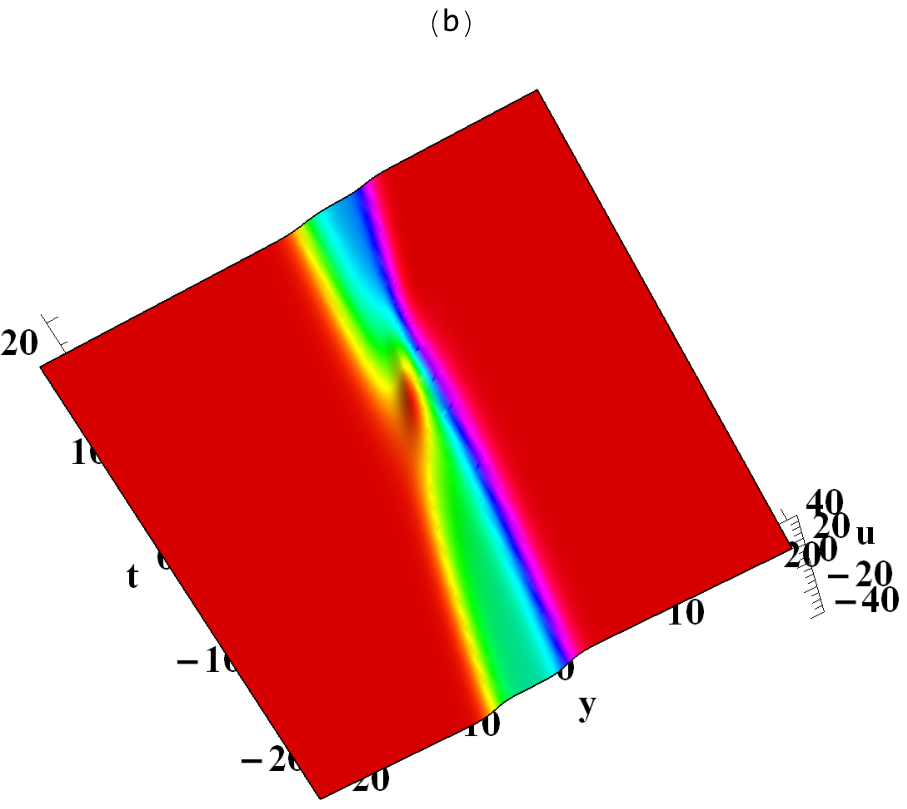}
\includegraphics[scale=0.45,bb=-300 250 10 10]{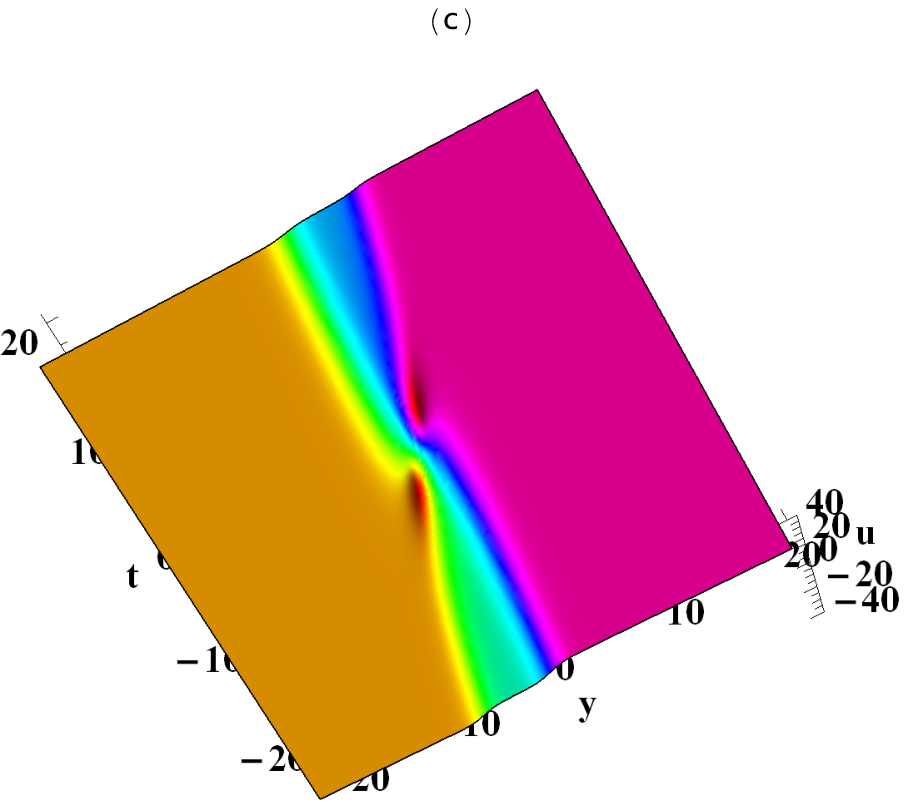}
\includegraphics[scale=0.45,bb=520 520 10 10]{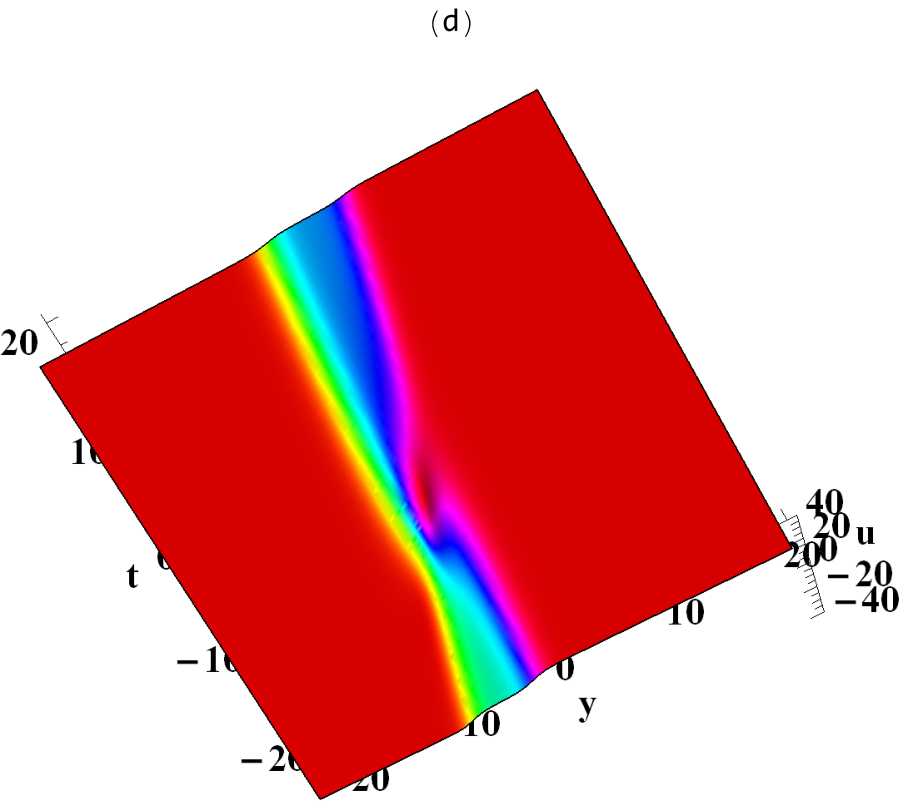}
\includegraphics[scale=0.45,bb=-400 520 10 10]{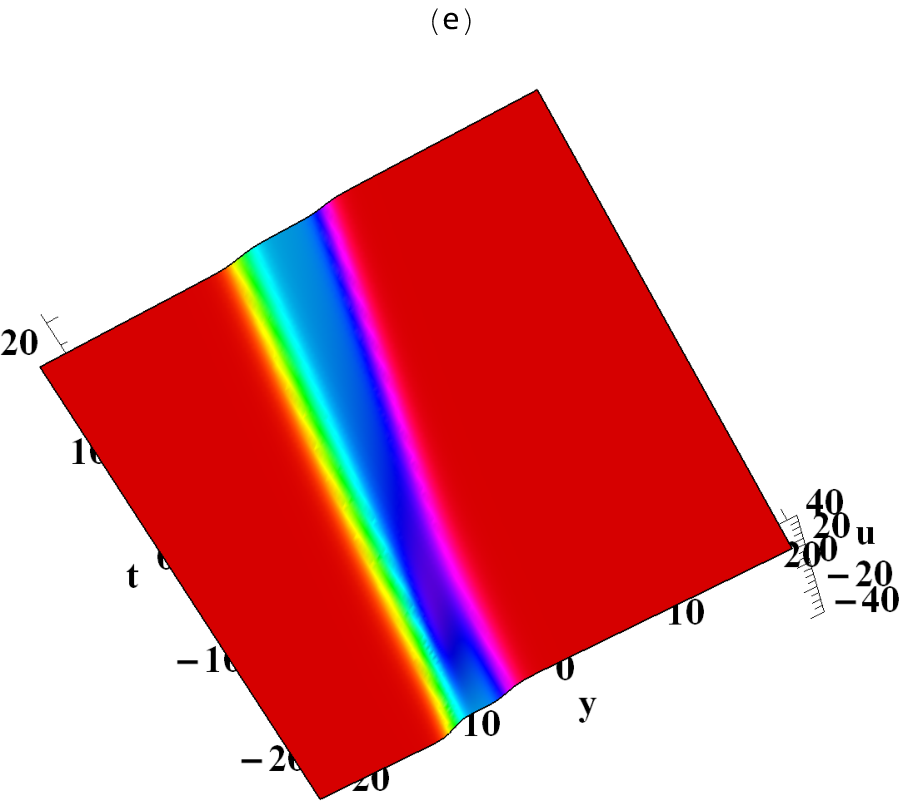}
\vspace{8cm}
\begin{tabbing}
\textbf{Fig.6}. Solution (20) in Eq. (19) with $b(t)=1$  when (a)
$x= -5$, (b) $x= -2$, (c) $x= 0$,\\ (d) $x= 2$, (e) $x= 5$.
\end{tabbing}

\includegraphics[scale=0.4,bb=80 300 10 10]{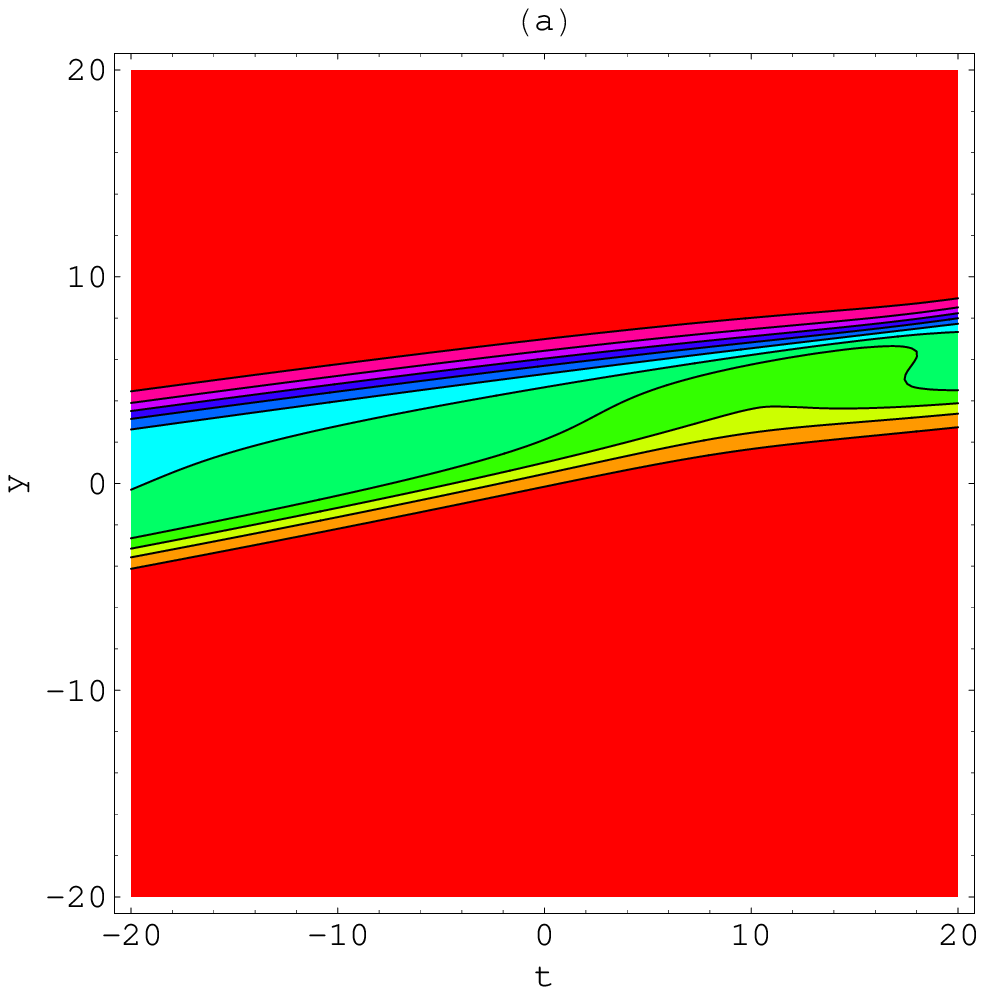}
\includegraphics[scale=0.4,bb=-335 300 10 10]{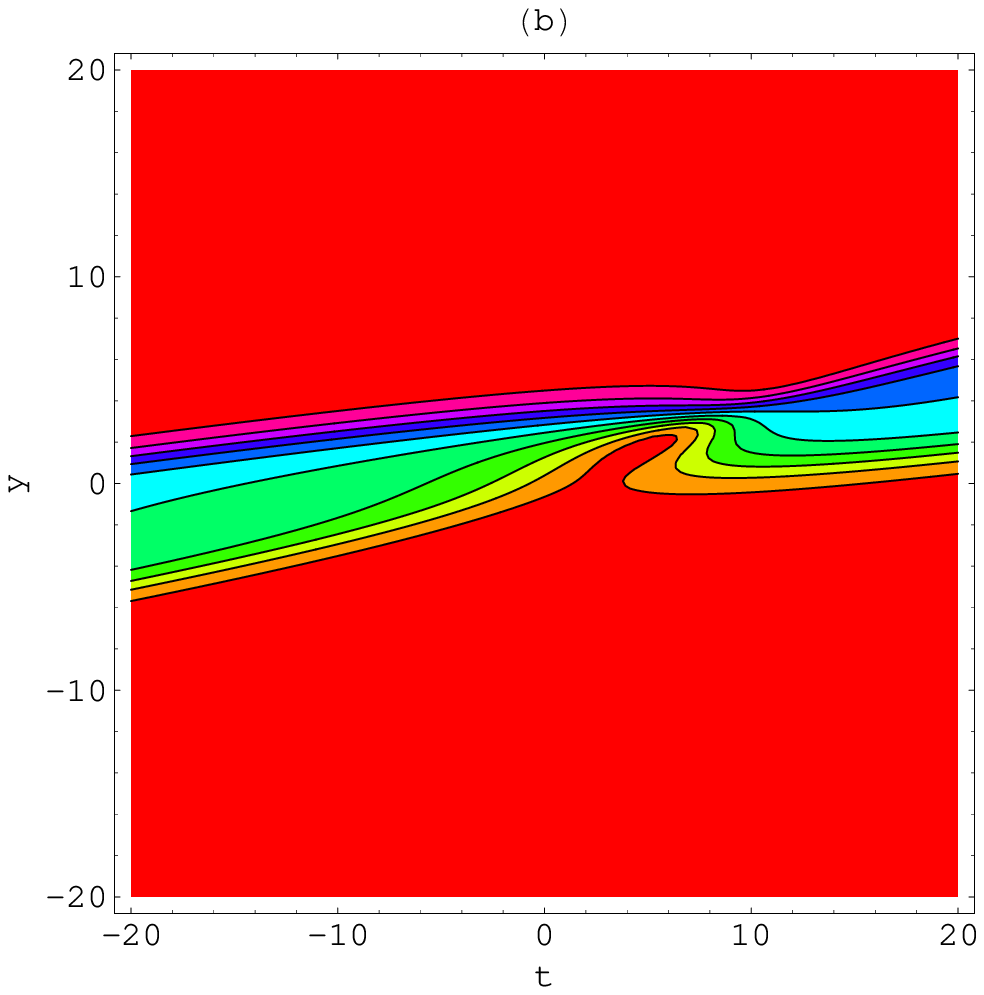}
\includegraphics[scale=0.4,bb=-340 300 10 10]{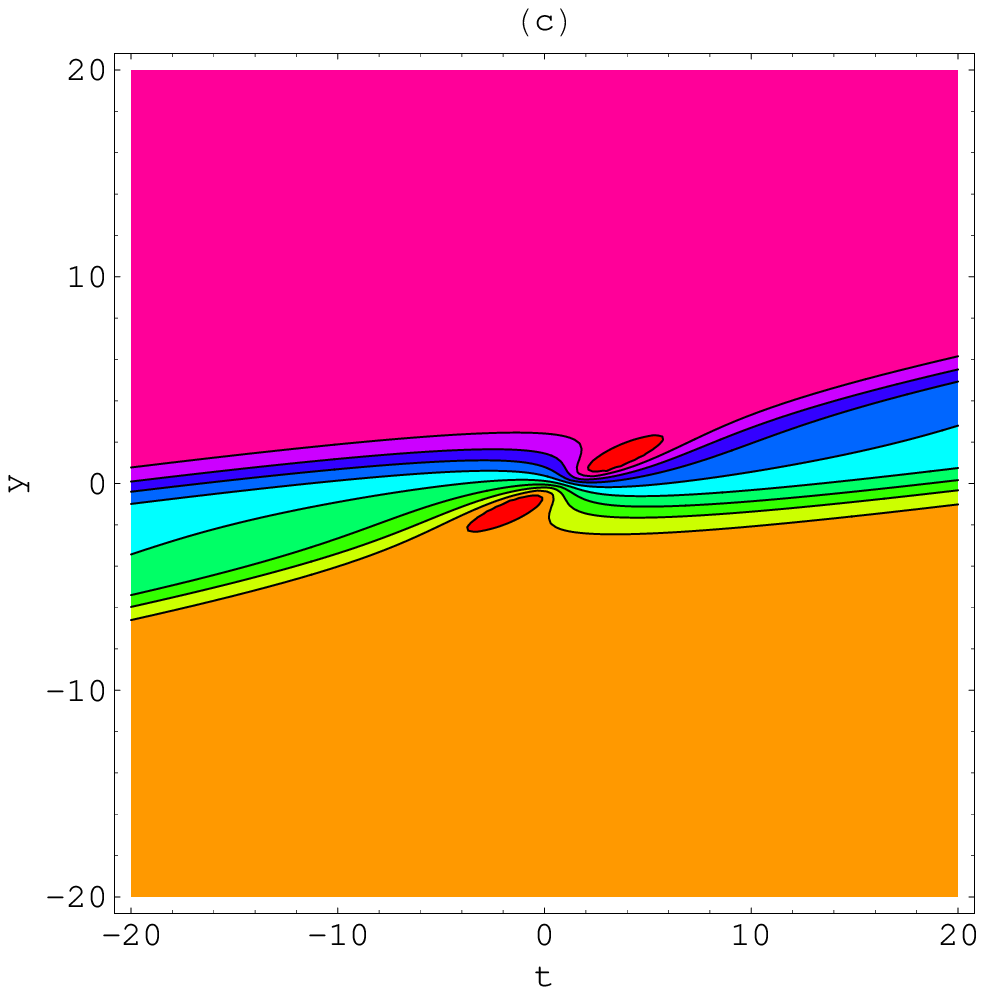}
\includegraphics[scale=0.4,bb=520 600 10 10]{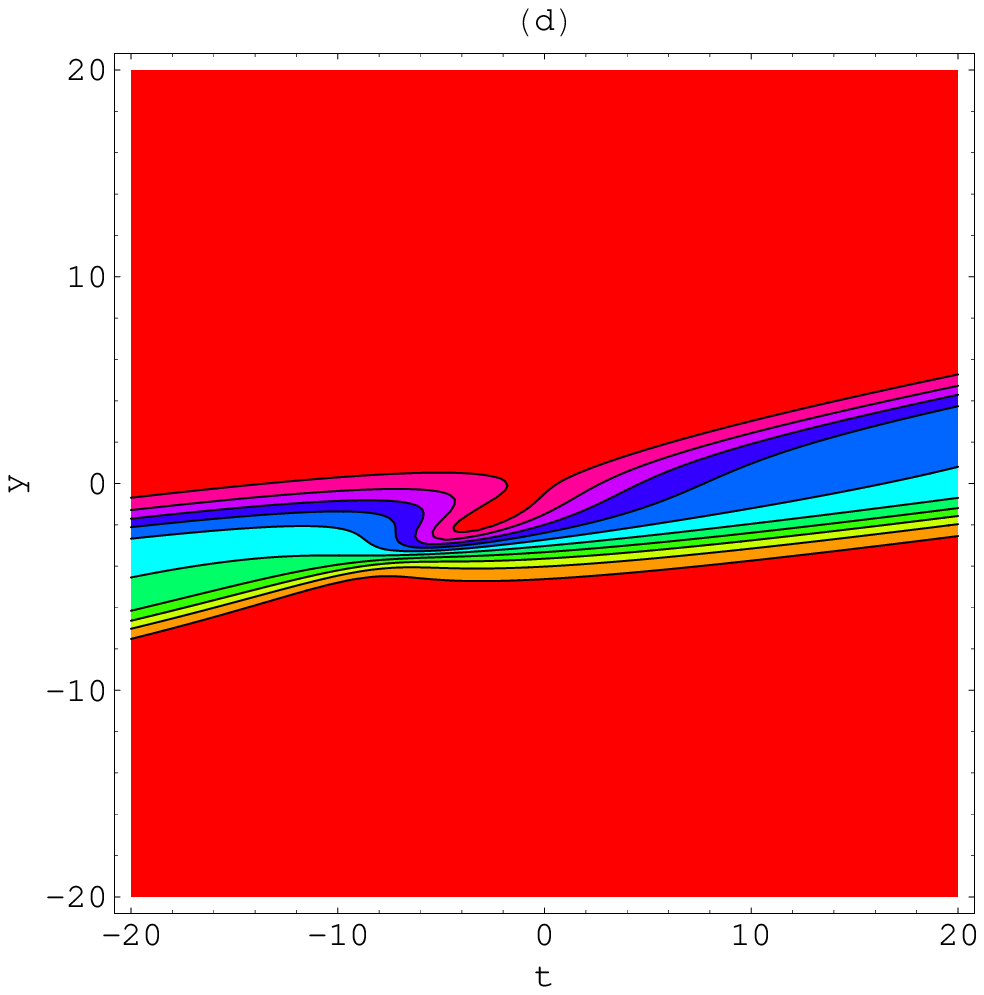}
\includegraphics[scale=0.4,bb=-400 600 10 10]{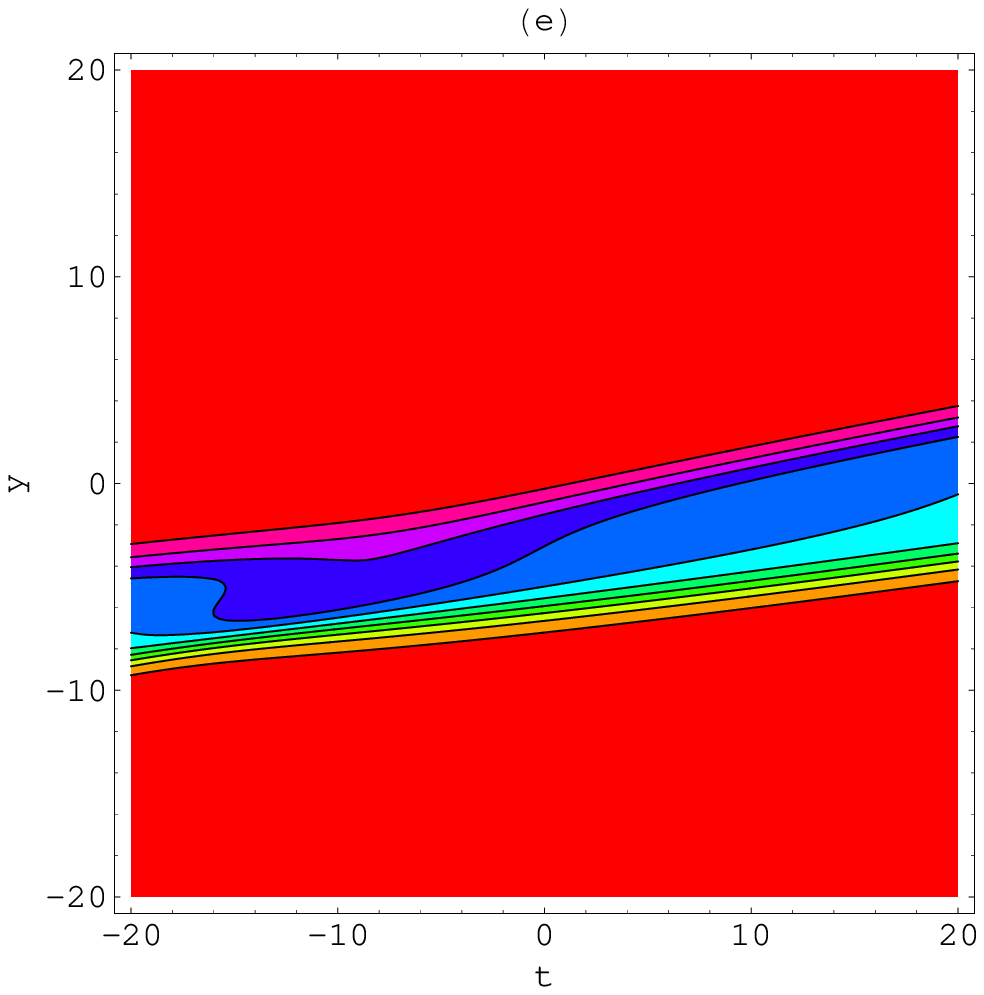}
\vspace{8cm}
\begin{tabbing}
\textbf{Fig.7}. The corresponding contour plots of Fig.6.
\end{tabbing}

\noindent {\bf 4. Lump-hyperbolic solution}\\

\quad Similar to the above lump-stripe solution,  the interaction
solutions between lump solution and the hyperbolic function (called
lump-hyperbolic solution) to Eq. (1) can be supposed as
follows  {\begin{eqnarray} \zeta&=&\vartheta _4(t)+x \vartheta _1+y \vartheta _2+z \vartheta _3,\nonumber\\
\varsigma&=&\vartheta _{9}(t)+x \vartheta _6+y \vartheta _7+z \vartheta _8,\nonumber\\
\xi&=& \zeta^2+\varsigma^2+\vartheta _{5}(t)+\alpha _{14}(t) \cosh
[\vartheta _{13}(t)+\vartheta _{10} x+\vartheta _{11} y+\vartheta
_{12} z].
\end{eqnarray}}Substituting Eq. (18) into Eq. (3) via the Mathematical
software, we have

 {\begin{eqnarray} \vartheta _7&=&-\frac{\vartheta
_1 \vartheta _2}{\vartheta _6},
   \vartheta_8=-\frac{\vartheta _1 \vartheta _3}{\vartheta _6}, \vartheta _5(t)=\vartheta _5,\nonumber\\
    a(t)&=&-\frac{4 \left(\vartheta _2^2+2 \vartheta _3 \vartheta _2+2 \vartheta _3^2\right) \vartheta _5 \left(\vartheta _1^2+\vartheta _6^2\right) b(t)}{3 \left(\vartheta
   _2^2+2 \vartheta _3 \vartheta _2+\vartheta _3^2+\vartheta _6^2\right) \vartheta _{10}^3 \vartheta _{11} \alpha _{14}(t){}^2}, \nonumber\\
\vartheta_4(t)&=& \eta _{18}+\int_1^t \frac{\left(\vartheta
_1-\vartheta _2-\vartheta _3\right) \left(\vartheta _3^2-\vartheta
_6^2\right) b(t)}{\left(\vartheta _2+\vartheta
_3\right){}^2+\vartheta _6^2} \, dt, \alpha_{14}(t)=\frac{\sqrt{2} \sqrt{\vartheta _5} \sqrt{\vartheta _1^2+\vartheta _6^2}}{\vartheta _{10}},\nonumber\\
\vartheta_{9}(t)&=&\eta _{19}+\int_1^t \frac{\left(\vartheta
_3^2-\vartheta _6^2\right) [\vartheta _6^2+\vartheta _1
\left(\vartheta _2+\vartheta _3\right)] b(t)}{\vartheta
   _6^3+\left(\vartheta _2+\vartheta _3\right){}^2 \vartheta _6} \,
   dt,\nonumber\\ \vartheta_{13}(t)&=&\eta _{20}-\int_1^t \frac{\vartheta _{11} \vartheta _{10}^3 a(t)+\left(\vartheta _{10}^2+\vartheta _{12}^2\right) b(t)}{\vartheta _{10}+\vartheta _{11}+\vartheta
   _{12}} \, dt,\nonumber\\
   \vartheta_{10}&=&\frac{\sqrt{3 \left(\vartheta _3^2-\vartheta _6^2\right) \vartheta _{11}^2+6 \left(\vartheta _3^2-\vartheta _6^2\right) \vartheta _{12} \vartheta _{11}-3
   \left(\vartheta _2^2+2 \vartheta _3 \vartheta _2+2 \vartheta _6^2\right) \vartheta _{12}^2}}{\sqrt{\vartheta _2^2
   +2 \vartheta _3 \vartheta _2+2 \vartheta _3^2}},\nonumber\\
   \vartheta_6&=&\frac{\sqrt{\left(\vartheta _3 \vartheta _{11}-\vartheta _2 \vartheta _{12}\right) [\vartheta _3 \left(2 \vartheta _2+\vartheta _3\right) \vartheta
   _{11}+\vartheta _2 \left(\vartheta _2+2 \vartheta _3\right) \vartheta _{12}]}}{\sqrt{\left(2 \vartheta _2+\vartheta _3\right) \vartheta _{11}^2+\left(3
   \vartheta _2+2 \vartheta _3\right) \vartheta _{12} \vartheta _{11}+2 \vartheta _2 \vartheta _{12}^2}},
\end{eqnarray}}where $\eta _{18}$, $\eta _{19}$ and $\eta _{20}$ are integral constants. Substituting  Eq. (19) into the
     transformation $u=\frac{6}{\rho}\,[ln\xi(x,y,z,t)]_x$, the lump-hyperbolic solution for Eq. (1) are derived as follows
{\begin{eqnarray} u&=& [6 [\vartheta _{10} \alpha _{14}(t) \sinh
[\vartheta _{13}(t)+x \vartheta _{10}+y \vartheta _{11}+z \vartheta
_{12}]+2 \vartheta _1
   [\vartheta _4(t)+x \vartheta _1+y \vartheta _2+z \vartheta _3]\nonumber\\&+&2 \vartheta _6 [\vartheta _9(t)+x \vartheta _6-\frac{y \vartheta _1 \vartheta
   _2}{\vartheta _6}-\frac{z \vartheta _1 \vartheta _3}{\vartheta _6}]]]/[\rho  [\alpha _{14}(t) \cosh [\vartheta _{13}(t)+x \vartheta _{10}+y
   \vartheta _{11}+z \vartheta _{12}]\nonumber\\&+&[\vartheta _4(t)+x \vartheta _1+y \vartheta _2+z \vartheta _3]{}^2+[\vartheta _9(t)+x \vartheta
   _6-\frac{y \vartheta _1 \vartheta _2}{\vartheta _6}-\frac{z \vartheta _1 \vartheta _3}{\vartheta _6}]{}^2+\vartheta _5]].
\end{eqnarray}}

 \quad Selecting the special values of the parameters as
{\begin{eqnarray} \vartheta_1=3, \vartheta_3=\vartheta_{11}=2,
 \rho=\vartheta_2=\vartheta_{12}=\vartheta_5=1,
\eta_{18}=\eta_{19}=\eta_{20}= z=0.
\end{eqnarray}}Substituting Eq. (21) into Eq. (20), the dynamical behaviors for solution (20) are shown in Fig. 6, Fig. 7 and Fig. 8.

\quad Fig. 6  reveals the interaction between lump and the
hyperbolic function with $b(t)=1$ in the $(t, y)$-plane. Fig. 7 lists
the corresponding contour plots of Fig. 6. when $x=-5, -2, 0, 2, 5$,
Lump disseminates and to get into hyperbolic function, and lump
turns into more  flourishing. Fig. 8  shows the dynamic properties
of solution (20) with $b(t)=t$.
\newpage
\includegraphics[scale=0.4,bb=-20 270 10 10]{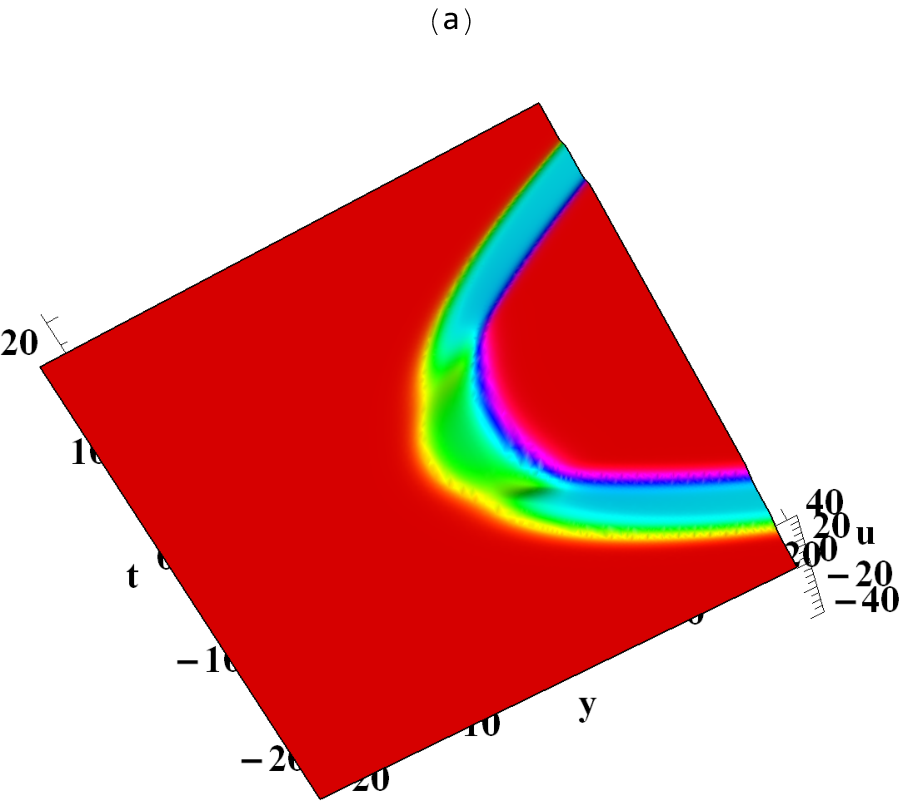}
\includegraphics[scale=0.4,bb=-355 270 10 10]{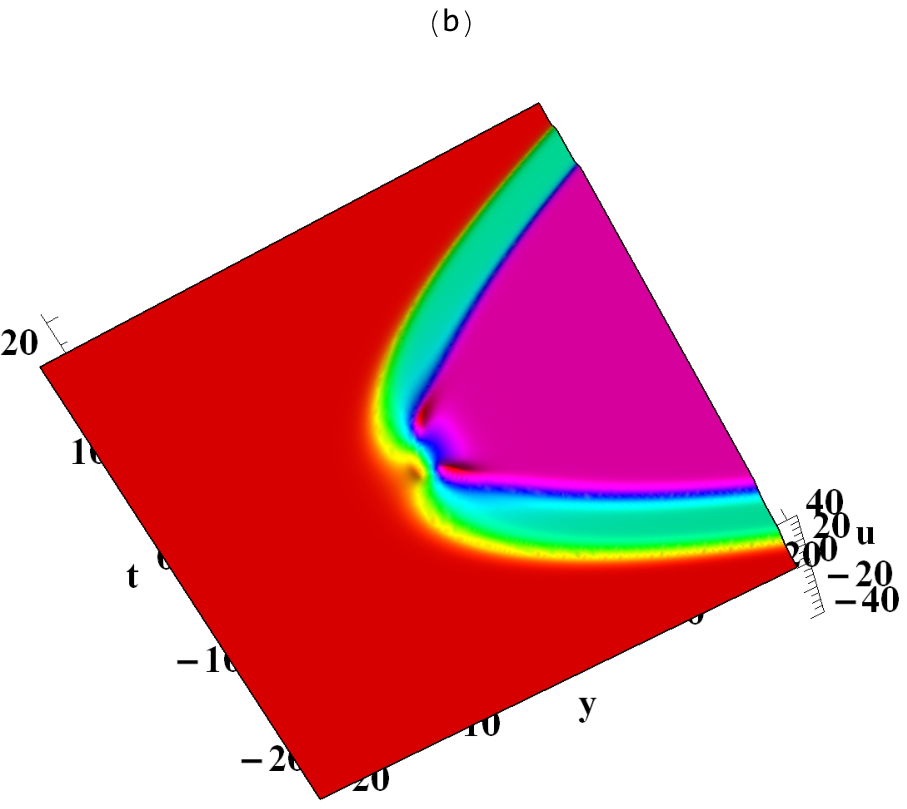}
\includegraphics[scale=0.4,bb=-360 270 10 10]{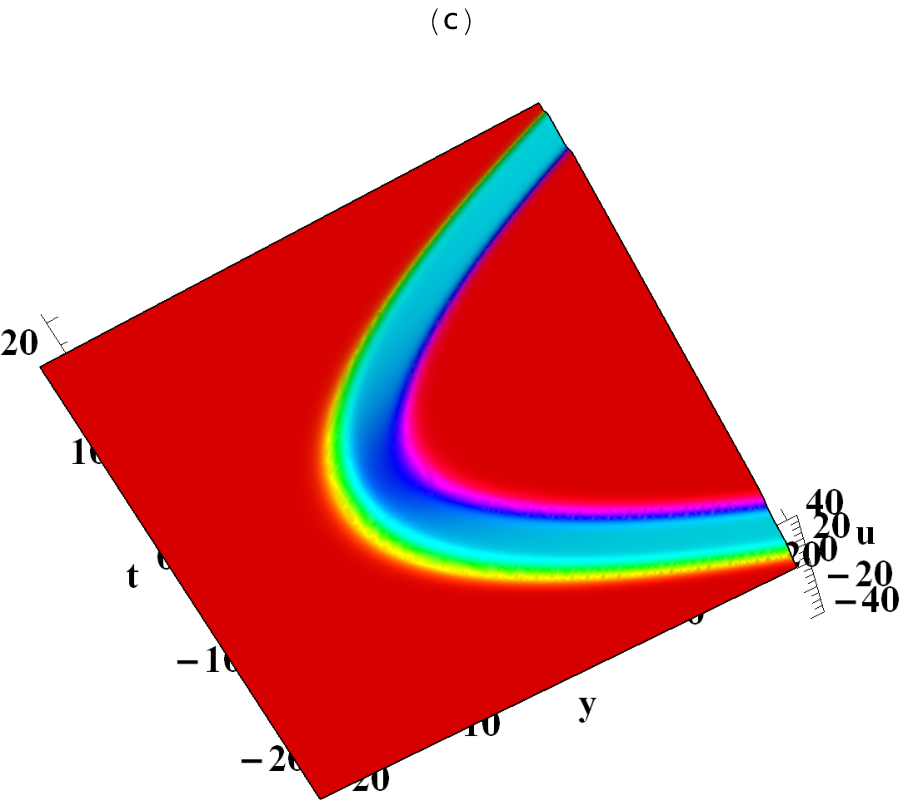}
\includegraphics[scale=0.35,bb=960 620 10 10]{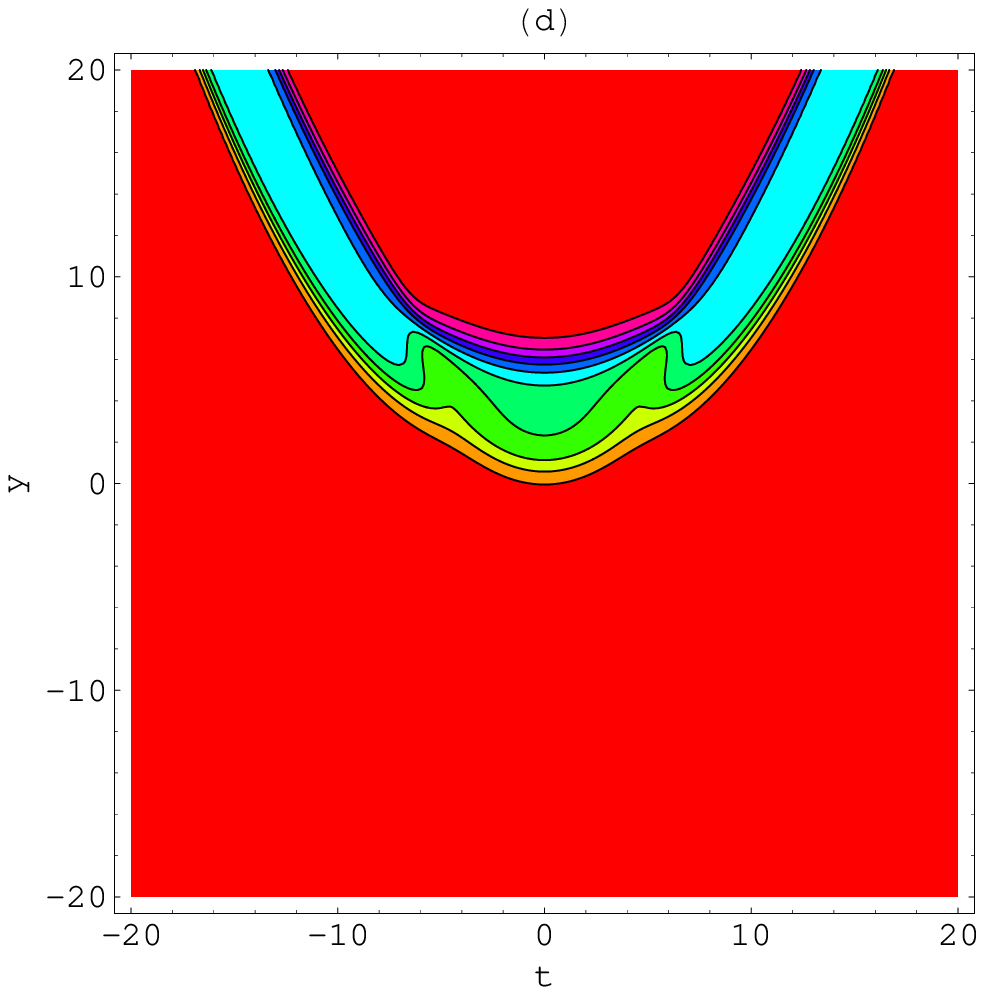}
\includegraphics[scale=0.35,bb=-405 620 10 10]{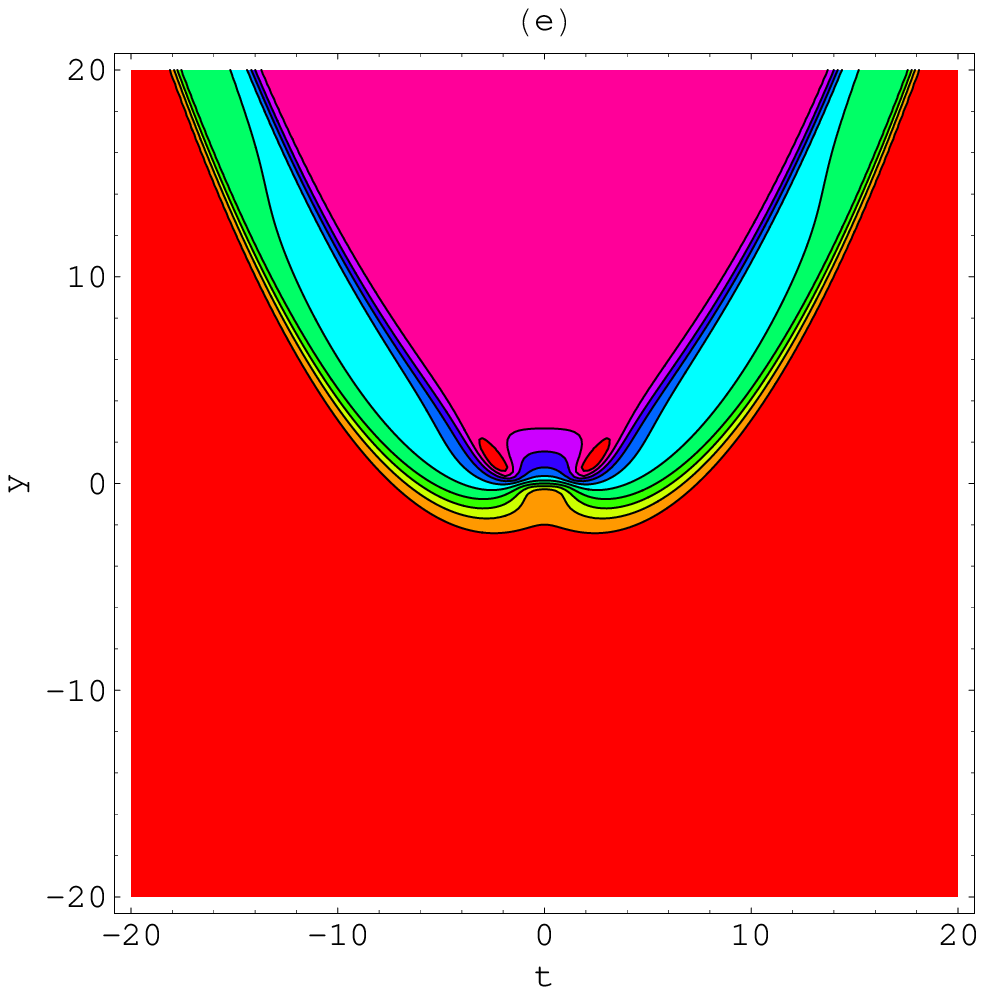}
\includegraphics[scale=0.35,bb=-400 620 10 10]{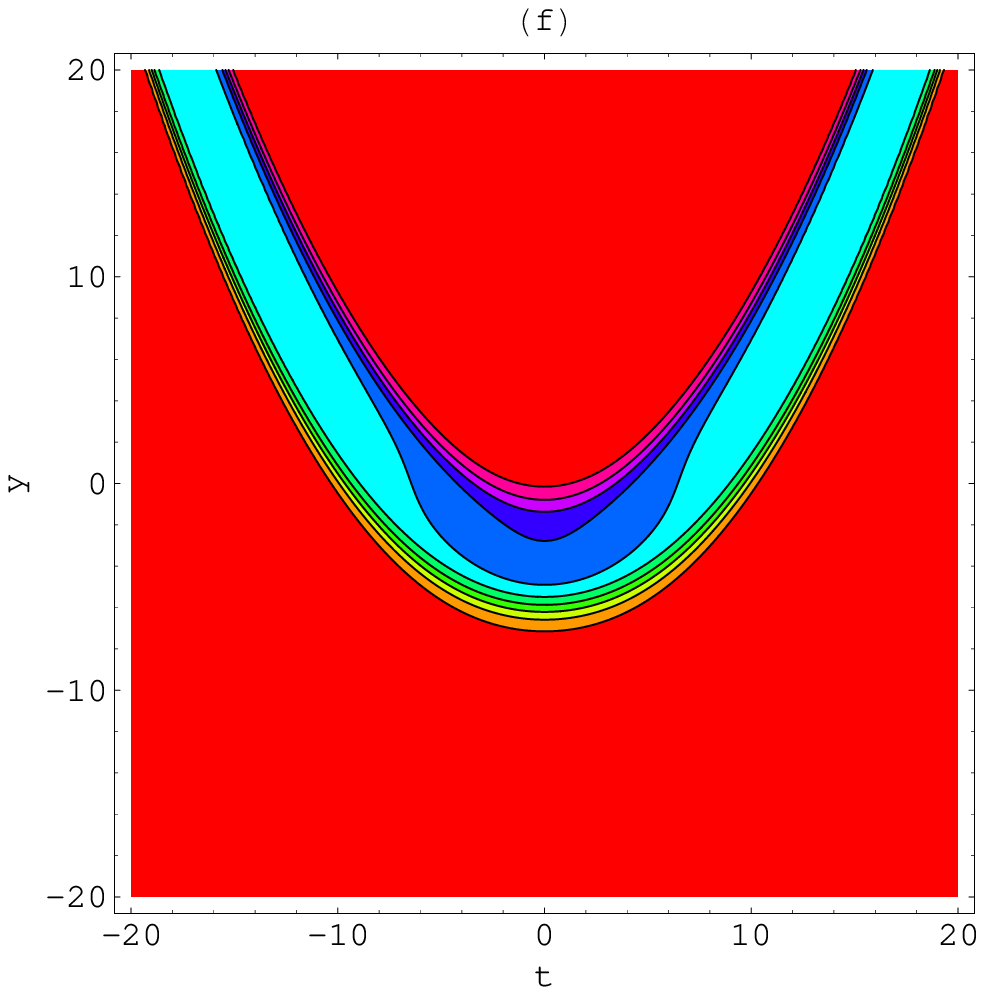}
\vspace{7.5cm}
\begin{tabbing}
\textbf{Fig.8}. Solution (20) in Eq. (19) with $b(t)=t$  when $x=
-5$  in (a) (d), $x= 0$ in (b) (e)\\ and $x = 5$ in (c) (f).\\
\end{tabbing}

\noindent {\bf 5. Conclusion}\\

\quad In this paper,  a generalized (3 + 1)-dimensional
variable-coefficient B-type KP equation is  investigated.  Abundant
lump solutions are presented based on Hirota's bilinear form and
symbolic computation. The interaction solutions  are obtained via
the lump solution and exponential function. The interaction
solutions between lump solution and hyperbolic function are also
studied.
Their dynamical behaviors are shown in Figs. 1-8.\\

\end{document}